\documentclass[preprint,prc,floats,tightenlines,showpacs,superscriptaddress,nofootinbib]{revtex4-1}
\usepackage{amsfonts}
\usepackage{amssymb}
\usepackage{amsmath}
\usepackage[dvips]{graphicx,color}
\usepackage{dcolumn}
\usepackage{epsfig}
\usepackage{bm}
\usepackage{url}

\newcommand{\ket}[1]{|{#1}\rangle}
\newcommand{\bra}[1]{\langle{#1}|}
\newcommand{\inp}[2]{\langle{#1}|{#2}\rangle}

\def\ohalf{\textstyle{\frac{1}{2}}}
\def\thalf{\textstyle{\frac{3}{2}}}
\def\fhalf{\textstyle{\frac{5}{2}}}
\def\shalf{\textstyle{\frac{7}{2}}}

\begin{document}

\title{Dynamical coupled-channels model of $K^- p$ reactions (II):
Extraction of $\Lambda^*$ and $\Sigma^*$ hyperon resonances}
\author{H. Kamano}
\affiliation{Research Center for Nuclear Physics, Osaka University, Ibaraki, Osaka 567-0047, Japan}
\author{S. X. Nakamura}
\affiliation{Department of Physics, Osaka University, Toyonaka, Osaka 560-0043, Japan}
\author{T.-S. H. Lee}
\affiliation{Physics Division, Argonne National Laboratory, Argonne, Illinois 60439, USA}
\author{T. Sato}
\affiliation{Department of Physics, Osaka University, Toyonaka, Osaka 560-0043, Japan}

\begin{abstract}
Resonance parameters (pole masses and residues) 
associated with the excited states of hyperons, $\Lambda^*$ and $\Sigma^*$,
are extracted within a dynamical coupled-channels model developed 
recently by us [Phys.~Rev.~C~{\bf 90}, 065204 (2014)]
through a comprehensive partial-wave analysis of 
the $K^- p \to \bar K N, \pi \Sigma, \pi \Lambda, \eta \Lambda, K\Xi$ data
up to invariant mass $W = 2.1$ GeV.
We confirm the existence of resonances corresponding to 
most, \textit{if not all}, of the four-star resonances rated by the Particle Data Group.
We also find several new resonances, and in particular
propose a possible existence of a new narrow $J^P=3/2^+$ $\Lambda$ resonance
that couples strongly to the $\eta \Lambda$ channel.
The $J^P=1/2^-$ $\Lambda$ resonances located below the $\bar K N$ threshold
are also discussed. 
Comparing our extracted pole masses with the ones from a recent analysis by 
the Kent State University group, 
some significant differences in the extracted resonance parameters are found, suggesting 
the need of more extensive and accurate data of $K^- p$ reactions including polarization 
observables to eliminate such an analysis dependence of the resonance parameters.
In addition, the determined large branching ratios of the decays of
high-mass resonances to the $\pi \Sigma^*$ and $\bar K^* N$ channels
also suggest the importance of the data of $2 \to 3$ reactions
such as $K^- p \to \pi\pi\Lambda$ and $K^- p \to \pi \bar K N$.
Experiments on measuring cross sections and polarization observables
of these fundamental reactions are highly desirable 
at hadron beam facilities such as J-PARC
for establishing the $\Lambda^*$ and $\Sigma^*$ spectrum.
\end{abstract}

\maketitle

\section{Introduction}

The spectra and structure of the excited baryons with light valence quarks ($u, d, s$) 
contain the information for understanding the non-perturbative aspects, confinements 
and chiral symmetry breaking, of Quantum Chromodynamics.
The excited baryons are unstable and couple with meson-baryon 
continuum states to form nucleon resonances ($N^*, \Delta^*$) with strangeness $S=0$ 
and hyperon resonances ($Y^* = \Lambda^*,\Sigma^*$) with $S=-1$.
Thus the extraction of these baryon resonances 
from the data of hadron-, photon-, and electron-induced meson-production reactions
has long been an important task in the hadron physics.
However, the hyperon resonances  are much less understood than the nucleon resonances.
This can be seen, for example, from the fact that only the Breit-Wigner masses and widths 
of the $\Lambda^*$ and $\Sigma^*$ resonances
were listed by the Particle Data Group (PDG) before 2012 
[exceptions are $\Sigma(1385)3/2^+$ and $\Lambda(1520)3/2^-$]~\cite{pdg2012}. 
In  contrast, the pole positions and residues of the $N^*$ and $\Delta^*$ resonances
have  been well determined by many analysis groups through detailed 
partial-wave analyses of $\pi N$ and $\gamma N$ reaction data.
To improve the situation, a first comprehensive and systematic partial-wave analysis 
of $K^- p$ reaction data to extract $Y^*$ resonance parameters defined by poles of 
scattering amplitudes was made in 2013 by the Kent State University (KSU) group using 
an energy-independent approach~\cite{zhang2013a}, 
and subsequently they extracted the pole masses of $Y^*$ resonances
by making an energy-dependent fit to their determined single-energy amplitudes~\cite{zhang2013b}.
Recently, we have also made an extensive partial-wave analysis of $K^-p$ reactions
within a dynamical-model approach~\cite{knlskp1}.
In this work, we present pole masses as well as residues of the $Y^*$ resonances
extracted from our amplitudes determined in Ref.~\cite{knlskp1}.

Our approach in Ref.~\cite{knlskp1} is based on a dynamical coupled-channels (DCC) formulation 
that was developed in Ref.~\cite{msl07} and applied extensively to 
the study of $N^*$ and $\Delta^*$ 
resonances~\cite{jlms07,jlmss08,jklmss09,kjlms09-1,sjklms10,
kjlms09-2,ssl2,knls10,knls13,kpi2pi13}.
Schematically, we solve the following coupled integral equations for the $T$-matrix elements
in each partial wave with strangeness $S=-1$,
\begin{equation}
T_{\beta,\alpha}(p_\beta,p_\alpha;W) = 
V_{\beta,\alpha}(p_\beta,p_\alpha;W)
+ \sum_{\delta} \int p^{2}d p  
V_{\beta,\delta}(p_\beta,p;W) 
G_{\delta}(p;W)
T_{\delta,\alpha}(p,p_\alpha;W) , 
\label{eq:teq}
\end{equation}
with
\begin{equation}
V_{\beta,\alpha}(p_\beta,p_\alpha;W)= 
v_{\beta,\alpha}(p_\beta,p_\alpha)
+ 
\sum_{Y^*}\frac{\Gamma^{\dagger}_{Y^*,\beta}(p_\beta)
 \Gamma_{Y^*,\alpha}(p_\alpha)} {W-M^0_{Y^*}} ,
\label{eq:veq}
\end{equation}
where $W$ is the invariant mass of the reaction;
the subscripts $\alpha,\beta,\delta$ denote 
the $\bar K N$, $\pi \Sigma$, $\pi \Lambda$, $\eta \Lambda$, and $K\Xi$ channels
as well as the quasi-two-body $\pi \Sigma^*$ and $\bar K^* N$ channels
that subsequently decay to the three-body $\pi \pi \Lambda$ and $\pi \bar K N$ channels;
$G_\delta$ is the Green's function of channel $\delta$; 
$M^0_{Y^*}$ is the mass of a bare excited hyperon state;
$v_{\alpha,\beta}$ is defined by hadron-exchange mechanisms; 
and the bare vertex interaction $\Gamma_{Y^*,\alpha}$ defines the $\alpha \to Y^*$ transition.
By fitting the data of both unpolarized and polarized observables of
the $K^- p \to \bar K N, \pi\Sigma, \pi\Lambda, \eta \Lambda, K\Xi$ reactions
over the wide energy range from the threshold to invariant mass $W = 2.1$~GeV,
we have constructed two models, called Model A and Model B~\cite{knlskp1}.
The partial-wave amplitudes and the $S$-wave threshold parameters 
such as scattering lengths and effective ranges were then extracted
from the constructed two models. The main objective of
this paper is to present the  
$\Lambda^*$ and $\Sigma^*$ resonance parameters extracted from these two models and
compare them with the results of the KSU analysis~\cite{zhang2013b}.

It is useful to mention here that the extracted resonance parameters are related to
the data through the mechanisms defined in the Hamiltonian of our model.
Thus it is possible to develop a theoretical understanding of the properties and structure
of the extracted resonances in our approach. 
Such a feature is not available in the KSU approach and
other similar approaches, in which the $K$ matrix or potential
are parametrized purely phenomenologically by using some polynomials, and so on.

In Sec.~\ref{sec:formulas}, we summarize notations, definitions, and formulas 
of the resonance parameters. 
In Sec.~\ref{sec:above}, the resonance parameters 
(mass spectrum, residues, and branching ratios, etc.)
extracted from our DCC models are presented for the $Y^*$ resonances 
located above the $\bar K N$ threshold, and 
the extracted mass spectra are compared with 
the one extracted from the KSU analysis.
We then give a prediction for the $S$-wave $\Lambda$ resonances 
located below the $\bar K N$ threshold in Sec.~\ref{sec:below}, 
which would be interesting in relation to $\Lambda(1405)$,
though it is a bit off the region of our current analysis.
Summary and discussions on future developments are given in Sec.~\ref{sec:summary}.

\section{Resonance parameters}
\label{sec:formulas}

\begin{table}
\caption{
The orbital angular momentum $(L)$ and total spin ($S$) of each $MB$ channel allowed in a given partial wave.
In the first column, partial waves are denoted with the conventional notation $l_{I2J}$ as well as ($I$,$J^P$).
\label{tab:qn}}
\begin{ruledtabular}
\begin{tabular}{ccccccccccc}
$l_{I2J}$ $(I,J^P)$ &\multicolumn{10}{c}{$(L,S)$ of the considered partial waves}\\
\cline{2-11}
                       &$\bar K N$  &$\pi \Sigma$&$\pi \Lambda$&$\eta\Lambda$&$K\Xi$     &\multicolumn{2}{c}{$\pi \Sigma^*$}&\multicolumn{3}{c}{$\bar K^* N$}      \\
                       &            &            &             &             &           &$(\pi \Sigma^*)_1$&$(\pi \Sigma^*)_2$ &$(\bar K^* N)_1$&$(\bar K^* N)_2$&$(\bar K^* N)_3$\\
\hline
$S_{01}$ $(0,\ohalf^-)$&($0,\ohalf$)&($0,\ohalf$)&--          &($0,\ohalf$)&($0,\ohalf$)&($2,\thalf$)& --         &($0,\ohalf$)&($2,\thalf$)& --         \\
$S_{11}$ $(1,\ohalf^-)$&($0,\ohalf$)&($0,\ohalf$)&($0,\ohalf$)&--          &($0,\ohalf$)&($2,\thalf$)&--          &($0,\ohalf$)&($2,\thalf$)& --         \\
$P_{01}$ $(0,\ohalf^+)$&($1,\ohalf$)&($1,\ohalf$)&--          &($1,\ohalf$)&($1,\ohalf$)&($1,\thalf$)&--          &($1,\ohalf$)&($1,\thalf$)& --         \\
$P_{03}$ $(0,\thalf^+)$&($1,\ohalf$)&($1,\ohalf$)&--          &($1,\ohalf$)&($1,\ohalf$)&($1,\thalf$)&($3,\thalf$)&($1,\ohalf$)&($1,\thalf$)&($3,\thalf$)\\
$P_{11}$ $(1,\ohalf^+)$&($1,\ohalf$)&($1,\ohalf$)&($1,\ohalf$)&--          &($1,\ohalf$)&($1,\thalf$)&--          &($1,\ohalf$)&($1,\thalf$)& --         \\
$P_{13}$ $(1,\thalf^+)$&($1,\ohalf$)&($1,\ohalf$)&($1,\ohalf$)&--          &($1,\ohalf$)&($1,\thalf$)&($3,\thalf$)&($1,\ohalf$)&($1,\thalf$)&($3,\thalf$)\\
$D_{03}$ $(0,\thalf^-)$&($2,\ohalf$)&($2,\ohalf$)&--          &($2,\ohalf$)&($2,\ohalf$)&($0,\thalf$)&($2,\thalf$)&($2,\ohalf$)&($0,\thalf$)&($2,\thalf$)\\
$D_{05}$ $(0,\fhalf^-)$&($2,\ohalf$)&($2,\ohalf$)&--          &($2,\ohalf$)&($2,\ohalf$)&($2,\thalf$)&($4,\thalf$)&($2,\ohalf$)&($2,\thalf$)&($4,\thalf$)\\
$D_{13}$ $(1,\thalf^-)$&($2,\ohalf$)&($2,\ohalf$)&($2,\ohalf$)&--          &($2,\ohalf$)&($0,\thalf$)&($2,\thalf$)&($2,\ohalf$)&($0,\thalf$)&($2,\thalf$)\\
$D_{15}$ $(1,\fhalf^-)$&($2,\ohalf$)&($2,\ohalf$)&($2,\ohalf$)&--          &($2,\ohalf$)&($2,\thalf$)&($4,\thalf$)&($2,\ohalf$)&($2,\thalf$)&($4,\thalf$)\\
$F_{05}$ $(0,\fhalf^+)$&($3,\ohalf$)&($3,\ohalf$)&--          &($3,\ohalf$)&($3,\ohalf$)&($1,\thalf$)&($3,\thalf$)&($3,\ohalf$)&($1,\thalf$)&($3,\thalf$)\\
$F_{07}$ $(0,\shalf^+)$&($3,\ohalf$)&($3,\ohalf$)&--          &($3,\ohalf$)&($3,\ohalf$)&($3,\thalf$)&($5,\thalf$)&($3,\ohalf$)&($3,\thalf$)&($5,\thalf$)\\
$F_{15}$ $(1,\fhalf^+)$&($3,\ohalf$)&($3,\ohalf$)&($3,\ohalf$)&--          &($3,\ohalf$)&($1,\thalf$)&($3,\thalf$)&($3,\ohalf$)&($1,\thalf$)&($3,\thalf$)\\
$F_{17}$ $(1,\shalf^+)$&($3,\ohalf$)&($3,\ohalf$)&($3,\ohalf$)&--          &($3,\ohalf$)&($3,\thalf$)&($5,\thalf$)&($3,\ohalf$)&($3,\thalf$)&($5,\thalf$)\\     
\end{tabular}
\end{ruledtabular}
\end{table}

Since the method for extracting the resonance parameters within the considered dynamical 
model has been explained
in detail in Refs.~\cite{ssl,ssl2,knls13}, here we just summarize
formulas that are needed for the presentations in this paper.

Consider the $MB \to M'B'$ reactions in the center-of-mass system, where $MB$ and $M'B'$ are the
initial and final meson-baryon states.
With the normalization $\inp{\vec{k}}{\vec{k}^{'}} =\delta(\vec{k}-\vec{k}^{'})$
for plane waves, the on-shell $S$ matrix elements
at the total scattering energy $W$ are given for each partial wave by
\begin{equation}
S_{M'B',MB}(W) = \delta_{M'B',MB} + 2iF_{M'B',MB}(W).
\label{eq:S}
\end{equation}
Here the index $MB$ ($M'B'$) also specifies quantum numbers associated with 
the channel $MB$ ($M'B'$), namely,
the orbital angular momentum ($L$), total spin ($S$), total angular momentum ($J$), 
parity ($P$), and isospin ($I$).
The values of these quantum numbers for the considered meson-baryon channels
are summarized in Table~\ref{tab:qn}.
The on-shell scattering amplitudes $F_{M'B',MB}(W)$ are related to
the $T$-matrix elements [Eq.~(\ref{eq:teq})] as
\begin{equation}
F_{M'B',MB}(W) = -
[\rho_{M'B'}(k_{M'B'}^{\text{on}};W)]^{1/2} 
T_{M'B',MB}(k_{M'B'}^{\text{on}}, k_{MB}^{\text{on}}; W) 
[\rho_{MB}(k_{MB}^{\text{on}};W)]^{1/2},
\label{eq:F-T}
\end{equation}
with
\begin{equation}
\rho_{MB}(k;W) = \pi \frac{kE_M(k) E_B(k)}{W},
\label{eq:rho}
\end{equation}
where $E_a(k) = \sqrt{m_a^2+k^2}$ is the energy of a particle $a$
with the mass $m_a$ and the three-momentum $\vec k$ ($k \equiv |\vec k|$).
For a given $W$, which can be complex, the on-shell momentum for the channel 
$MB$, $k_{MB}^{\mathrm{on}}$, is defined by 
\begin{equation}
W = E_M (k_{MB}^{\mathrm{on}}) + E_B (k_{MB}^{\mathrm{on}}).
\label{eq:on-shell-mom}
\end{equation}
The formulas and procedures to calculate the $T$-matrix elements within our DCC model
are fully explained in Ref.~\cite{knlskp1}, and thus we will not repeat them here.

As the energy $W$ approaches a pole position $M_R$ in the complex $W$ plane,
the scattering amplitudes take the following form,
\begin{equation}
F_{M'B',MB}(W\to M_R) = -\frac{R_{M'B',MB}}{W-M_R} + B_{M'B',MB} ,
\label{eq:F-pole}
\end{equation}
where $R_{M'B',MB}$ is the residue of $F_{M'B',MB}(W)$ at the resonance pole $M_R$, and
$B_{M'B',MB}$ is the ``background'' contribution. 
Both $R_{M'B',MB}$ and $B_{M'B',MB}$ are constant and in general complex.
The pole position ($M_R$) and the residue ($R_{M'B',MB}$) are fundamental quantities
that characterize the resonance.
In fact, within the resonance theory based on the Gamow vectors (see, e.g., Ref.~\cite{madrid}),
$M_R$ is equivalent to a complex energy eigenvalue of the \textit{total} Hamiltonian of the considered system
under the outgoing wave boundary conditions, and the (square-root of) residues
can be associated with the strength of the transition from the resonance 
to a scattering state of $MB$ and/or $M'B'$ channel.

Practically, within our approach
the value of $M_R$ for a resonance can be obtained as a solution
of the following equation with respect to $W$~\cite{ssl,ssl2,knls13}:
\begin{equation}
\mathrm{det}[D^{-1}(W)] = 0,
\label{eq:res1}
\end{equation}
with $D^{-1}(W)$ being the inverse of the dressed $Y^*$ resonance propagators.
It is defined by~\cite{knlskp1}
\begin{equation}
[D^{-1}(W)]_{n,m} = W \delta_{n,m} - [M_{Y^*}(W)]_{n,m}.
\label{eq:res2}
\end{equation}
The resonance mass matrix $M_{Y^*}(W)$ is given by
\begin{equation}
[M_{Y^*}(W)]_{n,m} = M_{Y_n^*}^0 \delta_{n,m} + [\Sigma_{Y^*}(W)]_{n,m},
\label{eq:res3}
\end{equation}
where $M_{Y_n^*}^0$ is the mass of the $n$th bare $Y^*$ state in a given partial wave,
and $\Sigma_{Y^*}(W)$ is the matrix for the $Y^*$ self energy~\cite{knlskp1}.
Solving Eq.~(\ref{eq:res1}) is nothing but searching for poles of the resonance propagator
in the complex $W$ plane.
The nonlinearity and multivaluedness of $\Sigma_{Y^*}(W)$ originating from 
the multichannel reaction dynamics make the relation between
bare states and physical resonances highly nontrivial.
In fact, as has been demonstrated in Ref.~\cite{sjklms10}, 
a naive one-to-one correspondence between bare states and physical resonances does not hold in
general within a multichannel reaction system.

It is noted~\cite{sjklms10} that Eqs.~(\ref{eq:res1})-(\ref{eq:res3}) give the exact 
resonance pole masses of the \textit{full} 
scattering amplitudes~(\ref{eq:F-T}) as far as the bare $Y^*$ state(s) is introduced 
for the considered partial wave.
Otherwise, one must search for resonance poles in the complex $W$ plane 
directly from the original full scattering amplitudes~(\ref{eq:F-T}).
Since at least one bare $Y^*$ state has been introduced for each partial wave
in our two models, Model A and Model B constructed in Ref.~\cite{knlskp1},
we just use Eq.~(\ref{eq:res1}) to search for resonance poles.

The residues $R_{M'B',MB}$ defined in Eq.~(\ref{eq:F-pole}) can be calculated by using the definition:
\begin{equation}
R_{M'B',MB} = \frac{1}{2\pi i}\oint_{C_{M_R}} dW [-F_{M'B',MB}(W)],
\label{eq:residue_def}
\end{equation}
where $C_{M_R}$ is an appropriate closed-path in the neighborhood of the point $W = M_R$,
circling $W = M_R$ in a counterclockwise manner.

As for the partial waves for which bare $Y^*$ state(s) is introduced, however,
$R_{M'B',MB}$ can also be calculated with~\cite{ssl,ssl2,knls13}
\begin{equation}
R_{M'B',MB} = 
[\rho_{M'B'}(k_{M'B'}^{\text{on}};M_R)]^{1/2} 
\bar \Gamma^R_{M'B',Y^*}(k_{M'B'}^{\text{on}};M_R)
\bar \Gamma^R_{Y^*,MB}(k_{MB}^{\text{on}};M_R)
[\rho_{MB}(k_{MB}^{\text{on}};M_R)]^{1/2}.
\label{eq:residue2}
\end{equation}
Here $\bar \Gamma^R_{MB,Y^*}(k;W)$ and $\bar \Gamma^R_{Y^*,MB}(k;W)$ 
are the dressed $Y^* \to MB$ and $MB \to Y^*$ vertices, respectively, given by
\begin{eqnarray}
\bar \Gamma^R_{MB,Y^*}(k;W) &=& \sum_n \chi_n \bar \Gamma_{MB,Y_n^*}(k;W),
\\
\bar \Gamma^R_{Y^*,MB}(k;W) &=& \sum_n \chi_n \bar \Gamma_{Y_n^*,MB}(k;W),
\end{eqnarray}
where 
$\bar \Gamma_{MB,Y_n^*}(k;W)$ and $\bar \Gamma_{Y_n^*,MB}(k;W)$, 
of which expressions are explicitly given in Ref.~\cite{knlskp1},
are the dressed vertices for the $n$th bare $Y^*$ state;
and the coefficient $\chi_n$ satisfies
\begin{equation}
[D(W)]_{n,m} = \frac{\chi_n\chi_m}{W-M_R} + (\textrm{regular~terms~at~} W=M_R),
\end{equation}
in the neighborhood of the point $W = M_R$, and
\begin{equation}
\sum_m [D^{-1}(M_R)]_{n,m}\chi_m = 0.
\end{equation}
We have confirmed that Eq.~(\ref{eq:residue2}) indeed gives exactly the same value 
as calculated from using Eq.~(\ref{eq:residue_def}).

It should be emphasized here that the coefficient 
$\chi_n$ represents the $n$th bare-state
component of the fully dressed $Y^*$ resonance. 
In other words, it indicates the meson-baryon contents of a resonance within 
the dynamical reaction models~\cite{ttm}.
For example, for the case that one bare state is contained, 
the coefficient $\chi\equiv\chi_1$ is given explicitly as
\begin{equation}
\chi = \left( 1 - \left.\frac{d \Sigma_{Y^*}(W)}{d W}\right|_{W=M_R} \right)^{-1/2}.
\end{equation}
This is nothing but the square root of the (complex) wave function renormalization constant $Z$ for the bare state~\cite{ttm}.

\section{$\Lambda^*$ and $\Sigma^*$ resonances above the $\bar K N$ threshold}
\label{sec:above}

With the formulas described in Sec.~\ref{sec:formulas} and 
the analytic continuation method developed in Refs.~\cite{ssl,ssl2},
we have extracted the parameters of the $\Lambda^*$ and $\Sigma^*$ resonances
from Model A and Model B constructed in Ref.~\cite{knlskp1}.
In this section, we present the results for the resonances found in the energy
region above the $\bar K N$ threshold.

\subsection{Resonance masses}
\label{sec:res-mass}

\begin{table}
\caption{\label{tab:res-pole}
Extracted complex pole masses ($M_R$) for the $\Lambda^*$ and $\Sigma^*$ resonances
found in the energy region above the $\bar K N$ threshold.
The masses are listed as $\bm{(} {\rm Re}(M_R), -{\rm Im}(M_R)\bm{)}$
together with their deduced uncertainties.
The resonance poles are searched in the complex $W$ region with 
$m_{\bar K} + m_N \leq {\rm Re}(W) \leq 2.1$ GeV and $0\leq -{\rm Im}(W) \leq 0.2$ GeV,
and all of the resonances listed are located in the complex $W$ Riemann surface
nearest to the physical real $W$ axis. 
}
\begin{ruledtabular}
\begin{tabular}{lccc}
            & & \multicolumn{2}{c}{$M_R$ (MeV)}  \\
\cline{3-4}
&$J^P(l_{I2J})$ & Model A & Model B \\
\hline
$\Lambda$-baryons
&$1/2^-(S_{01})$&(1669$^{+3}_{-8}$, 9$^{+9}_{-1}$)&(1512$^{+1}_{-1}$,185$^{+1}_{-2}$)\\
&              &          &(1667$^{+1}_{-2}$,12$^{+3}_{-1}$)\\
&$1/2^+(P_{01})$&(1544$^{+3}_{-3}$, 56$^{+6}_{-1}$)&(1548$^{+5}_{-6}$, 82$^{+7}_{-7}$)\\
&              &(2097$^{+40}_{-1}$, 83$^{+32}_{-6}$)&(1841$^{+3}_{-4}$, 31$^{+3}_{-2}$)\\
&$3/2^+(P_{03})$&(1859$^{+5}_{-7}$,  56$^{+10}_{-2}$)&(1671$^{+2}_{-8}$, 5$^{+11}_{-2}$)\\
&$3/2^-(D_{03})$&(1517$^{+4}_{-4}$,  8$^{+5}_{-4}$)&(1517$^{+4}_{-3}$, 8$^{+6}_{-6}$)\\
&              &(1697$^{+6}_{-6}$, 33$^{+7}_{-7}$)&(1697$^{+6}_{-5}$, 37$^{+7}_{-7}$)\\
&$5/2^-(D_{05})$&(1766$^{+37}_{-34}$,106$^{+47}_{-31}$)&(1924$^{+52}_{-24}$, 45$^{+57}_{-17}$)\\
&              &(1899$^{+35}_{-37}$, 40$^{+50}_{-17}$)&          \\
&$5/2^+(F_{05})$&(1824$^{+2}_{-1}$, 39$^{+1}_{-1}$)&(1821$^{+1}_{-1}$, 32$^{+1}_{-1}$)\\
&$7/2^+(F_{07})$&(1757, 73)&(2041$^{+80}_{-82}$,119$^{+57}_{-17}$)\\
\\
$\Sigma$-baryons
&$1/2^-(S_{11})$&(1704$^{+3}_{-6}$, 43$^{+7}_{-2}$)&(1551$^{+2}_{-9}$,188$^{+6}_{-1}$)\\
&              &          &(1940$^{+2}_{-2}$, 86$^{+2}_{-2}$)\\
&$1/2^+(P_{11})$&(1547$^{+111}_{-59}$, 92$^{+43}_{-39}$)&(1457$^{+5}_{-1}$, 39$^{+1}_{-4}$)\\
&              &(1706$^{+67}_{-60}$, 51$^{+79}_{-42}$)&(1605$^{+2}_{-4}$, 96$^{+1}_{-5}$)\\
&              &          &(2014$^{+6}_{-13}$, 70$^{+14}_{-1}$)\\
&$3/2^-(D_{13})$&(1607$^{+13}_{-11}$,126$^{+15}_{-9}$)&(1492$^{+4}_{-7}$, 69$^{+4}_{-7}$)\\
&              &(1669$^{+7}_{-7}$, 32$^{+5}_{-7}$)&(1672$^{+5}_{-10}$, 33$^{+3}_{-3}$)\\
&$5/2^-(D_{15})$&(1767$^{+2}_{-2}$, 64$^{+2}_{-1}$)&(1765$^{+2}_{-1}$, 64$^{+3}_{-1}$)\\
&$5/2^+(F_{15})$&(1890$^{+3}_{-2}$, 49$^{+2}_{-3}$)&(1695$^{+20}_{-77}$, 97$^{+50}_{-44}$)\\
&$7/2^+(F_{17})$&(2025$^{+10}_{-5}$, 65$^{+3}_{-12}$)&(2014$^{+12}_{-1}$,103$^{+3}_{-9}$)
\end{tabular}
\end{ruledtabular}
\end{table}
The resonance masses (pole positions),  $M_R$, are the solutions of Eq.~(\ref{eq:res1}) 
on the complex $W$ plane. 
In general, the physical observables are less influenced by
the resonances with very large widths, and thus
the information for such resonances extracted from fitting the data are less reliable.
Therefore, following our previous study of $N^*$ and $\Delta^*$ resonances~\cite{knls13}, 
we examine only the resonances with  total width less than 400~MeV 
[the total width is defined as $\Gamma^{\rm tot}= -2{\rm Im}(M_R)$].
We also do not search for  resonances with ${\rm Re}(M_R) > 2.1$~GeV.
With these criteria, 18 (20) resonances with the spin-parity 
$J^P=1/2^{\pm},3/2^{\pm},5/2^{\pm},7/2^+$ are extracted within the
Model A (Model B) in the energy region above the $\bar K N$ threshold.
All of these resonances are located in the Riemann surface which is
\textit{nearest} to the physical real $W$ axis.
The extracted  resonance masses $M_R$ are listed in Table~\ref{tab:res-pole}.

Because of the ``incompleteness'' of the current database of the $K^- p$ reactions,
as discussed in Ref.~\cite{knlskp1}, there are expected to be different solutions of
partial-wave analysis with similar $\chi^2$ minima.
We indicate in Table~\ref{tab:res-pole} our estimates of the uncertainties 
of the extracted resonance masses originating from such ``indistinguishable'' solutions
within the available $K^- p$ reaction data.
Similar uncertainties also occur in the analyses of the $\pi N$ and $\gamma N$ reactions, 
as discussed, for example, in Ref.~\cite{bg2012}.
In principle, the uncertainties in a dynamical coupled-channels analysis, such as this work,
should be evaluated by varying all parameters of the starting Hamiltonian
simultaneously in wide ranges around
the values determined in the $\chi^2$ fits.
Such a procedure is however practically not feasible  since 
solving the coupled-channels integral 
equations~(\ref{eq:teq}) and~(\ref{eq:veq}) is rather time consuming and the parameter 
space of the constructed models is rather large.

Instead, we take a more tractable procedure described as follows.
For each partial wave listed in Table~\ref{tab:qn}, additional parameters,
$\delta M_{Y_n^*}$, are added to the diagonal elements of the mass
matrix [Eq.(\ref{eq:res3})]:
\begin{equation}
[M_{Y^*}(W)]_{n,m} \to [M_{Y^*}(W)]_{n,m}  + \delta M_{Y^*_n} \delta_{n,m},
\end{equation}
where $\delta M_{Y_n^*}$ are taken to be complex.
We then refit the $K^- p$ reaction data listed in Table II
of Ref.~\cite{knlskp1} by choosing randomly the initial values of
$\delta M_{Y_n^*}$.
By keeping the model parameters in the Hamiltonian fixed and
varying only the additional $\delta M_{Y_n^*}$ parameters
with a gradient minimization procedure, we obtain a set of $\delta M_{Y_n^*}$
values for the chosen initial values.
This minimization procedure is repeated about $10^5$ times for a wide
range of initial $\delta M_{Y_n^*}$ values.
We then pick up the solutions that give almost the same $\chi^2$
values as the original one ($\chi^2_{\textrm{org}}$) from Model A or B by setting
the condition $|(\chi^2-\chi^2_{\textrm{org}})/\chi^2_{\textrm{org}} | \leq 1\%$.
About 20$\%$ of the solutions meet this condition.
Note that this procedure of determining the range of allowed 
$\delta M_{Y_n^*}$ values, i.e., Monte Carlo sampling combined with gradient minimization,
is motivated by the one taken in Ref.~\cite{shkl11}
in determining the multipole amplitudes of the $\gamma p \to K^+ \Lambda$ reaction.
The uncertainties of the resonance masses are then determined from the range of pole values 
found by solving Eq.~(\ref{eq:res1}) in which $\delta M_{Y_n^*}$ is varied over the allowed range.
The resulting uncertainties are listed in Table~\ref{tab:res-pole}.
Overall, the magnitude of our estimated uncertainties are consistent 
with the one listed in PDG~\cite{pdg2014}\footnote{
A direct comparison of our uncertainties with those listed in PDG
may not be well-justified because the former is associated with the pole masses
whereas the latter is associated with the Breit-Wigner masses and widths.}.
One exception is the mass for the $J^P = 7/2^+$ $\Lambda$ resonance in Model A, 
for which the uncertainty is not assigned because it is too large to be meaningful.
Several resonances, e.g., $J^P=3/2^-$ $\Lambda$ resonances,
appear in both Models A and B to have almost the same central value and uncertainty for $M_R$.
Resonances that are found in only either of Model A or B basically have large uncertainty in their masses.
However, some exceptions also exist, for example, the first $J^P=1/2^-$ $\Lambda$ resonance in Model B,
for which the counterpart is not found in Model A but the uncertainty for its mass is rather small.
The existence of such exceptions implies that the dynamical contents of our two models 
are rather different from each other, yet they are hard to be distinguished with 
the current $K^- p$ reaction data included in our fits due to its ``incompleteness'' mentioned above 
and explained in Ref.~\cite{knlskp1}.

\begin{figure}
\includegraphics[clip,width=0.8\textwidth]{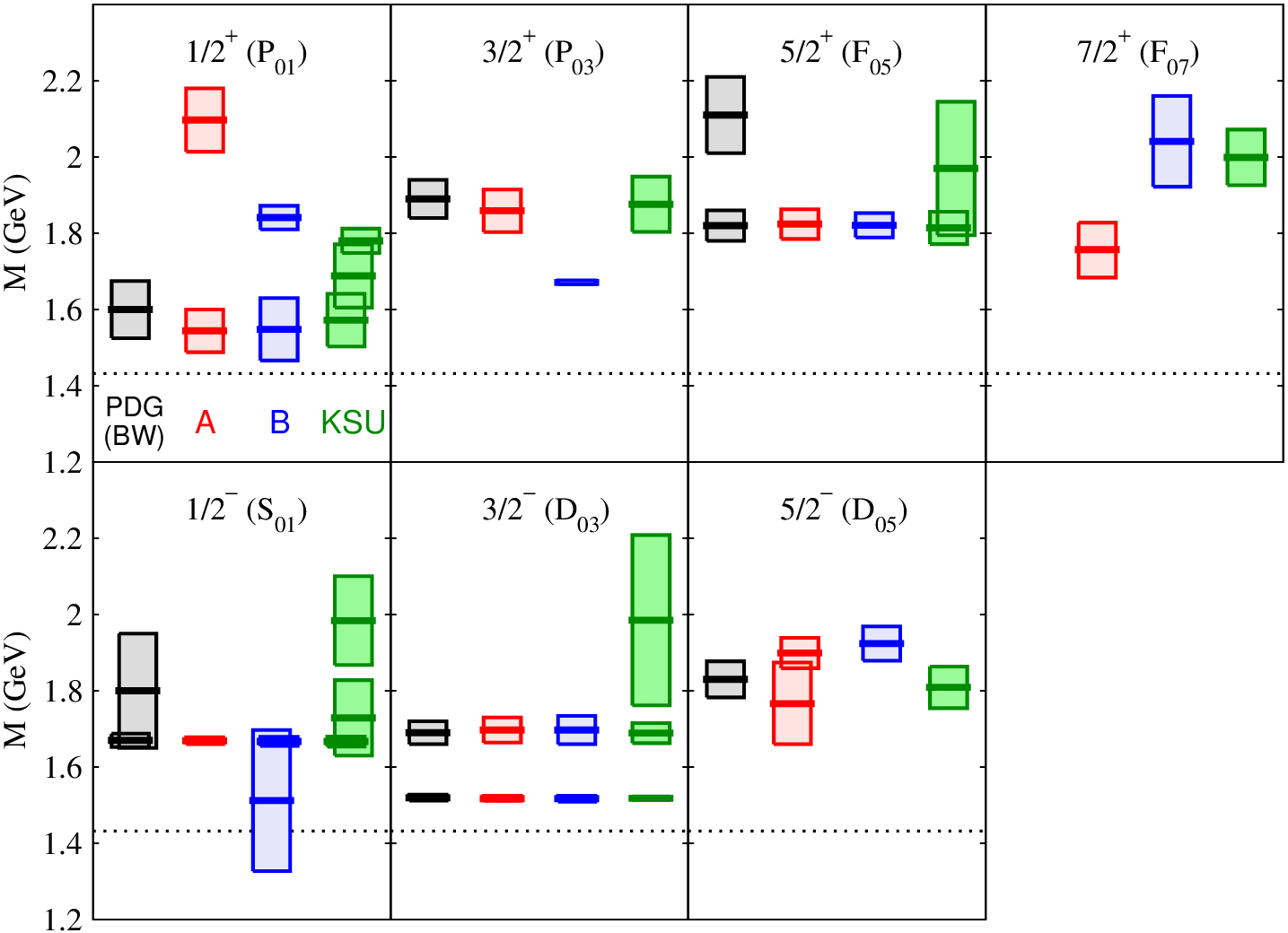}
\caption{(Color online)
Mass spectrum of $\Lambda^*$ resonances above the $\bar K N$ threshold.
For each $\Lambda^*$, $\mathrm{Re}(M_R)$ together with 
the $\mathrm{Re}(M_R)\pm[-\mathrm{Im}(M_R)]$ band is plotted, where the
length of the band, $-2\mathrm{Im}(M_R)$, corresponds to the total width of the resonance.
The spin and parity of the resonances are denoted as $J^P$ with $P=\pm$,
and also specified by the quantum number ($l_{I2J}$)
of the associated $\bar K N$ partial-wave amplitudes.
The horizontal dotted lines represent the $\bar K N$ threshold.
The results from Model A and Model B constructed in Ref.~\cite{knlskp1} 
are compared with the ones from the KSU analysis~\cite{zhang2013b}.
The so-called Breit-Wigner masses and widths
of the four- and three-star resonances rated by PDG~\cite{pdg2014} are also presented. 
}
\label{fig:l-spectrum}
\end{figure}
\begin{figure}
\includegraphics[clip,width=0.8\textwidth]{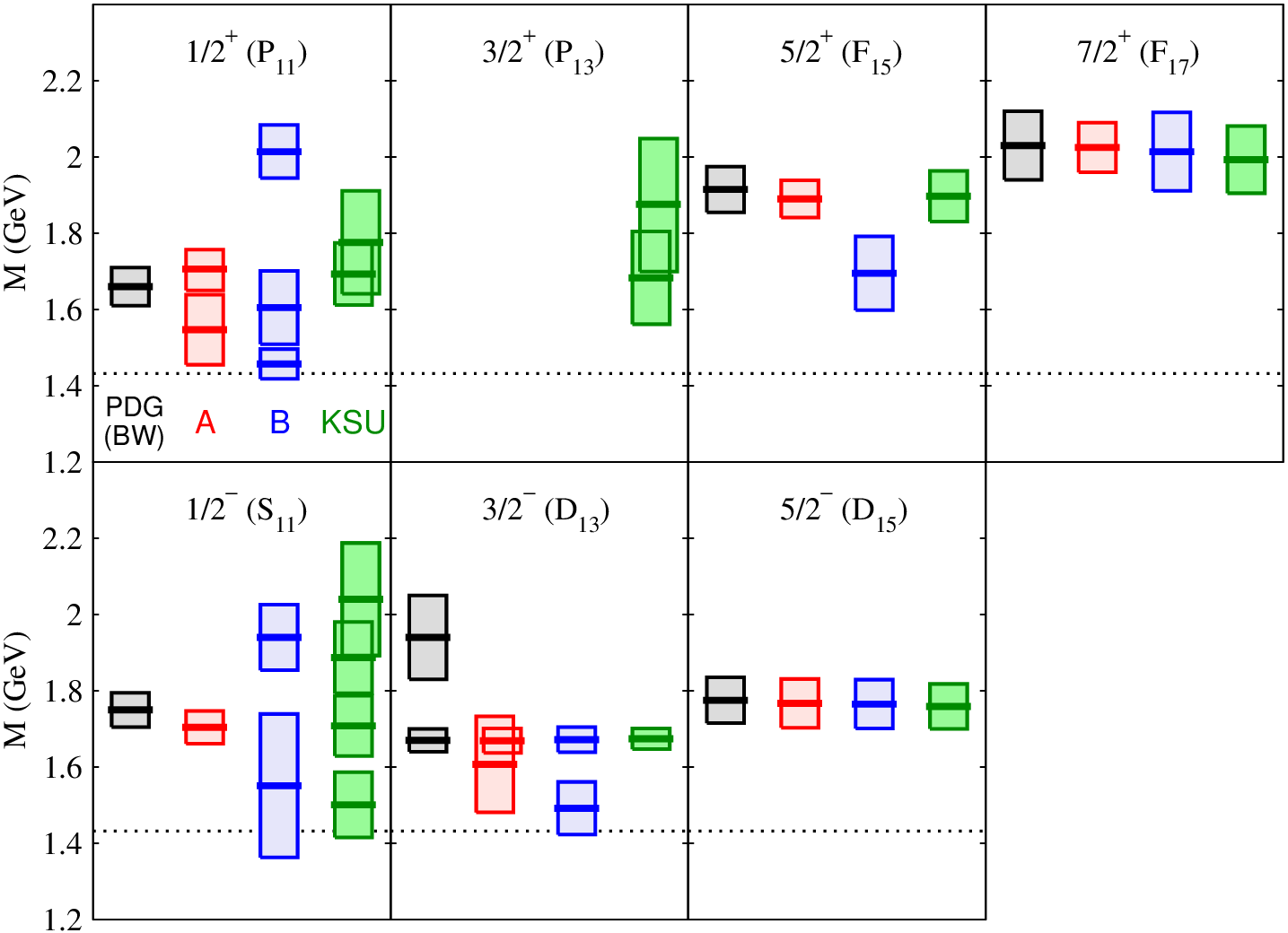}
\caption{(Color online)
Mass spectrum of $\Sigma^*$ resonances above the $\bar K N$ threshold.
See the caption of Fig.~\ref{fig:l-spectrum} for the description of the figure.
}
\label{fig:s-spectrum}
\end{figure}
The  $\Lambda^*$ and $\Sigma^*$ mass spectra extracted
from Models A and B  
are compared with the results from the KSU analysis~\cite{zhang2013b}
in Figs.~\ref{fig:l-spectrum} and~\ref{fig:s-spectrum}, respectively.
In the same figures, we also indicate the mass spectra of the 
four- and three-star resonances assigned by PDG~\cite{pdg2014}.
It should be noted that the mass spectra
listed by PDG~\cite{pdg2014} are evaluated using the  masses and widths from the
the Breit-Wigner parametrization of the scattering amplitudes.
It is now well recognized that the resonance parameters obtained with
the Breit-Wigner parametrizations
are not trivially related to the ones determined at the resonance pole positions.
Thus the PDG values given in Figs.~\ref{fig:l-spectrum} and~\ref{fig:s-spectrum} 
are just for an additional reference in assessing the model dependence of the analyses.

We see in Figs.~\ref{fig:l-spectrum} and~\ref{fig:s-spectrum} that
the results from our two models and the KSU analysis show an excellent agreement for several
resonances.
However, large discrepancies are also seen between the three results, and those 
need to be resolved.
This of course reflects the fact that the existing $K^- p$ reaction 
data are not sufficient to constrain the mass spectrum of $Y^*$ resonances
(see Ref.~\cite{knlskp1} for the current situation of the world data for 
the $K^- p$ reactions). 
More extensive and accurate data including polarization observables are 
highly desirable to get convergent results.

It is interesting to see that our two models and the KSU analysis have  low-lying $\Sigma^*$ resonances 
with $\mathrm{Re}(M_R) < 1.6$ GeV in $S_{11}$, $P_{11}$, and $D_{13}$.
They may correspond to one- and two-star resonances
assigned by PDG (but are not indicated in Fig.~\ref{fig:s-spectrum}).
To establish such low-lying resonances, more data near the $\bar K N$ threshold are definitely needed.

\begin{figure}
\includegraphics[clip,width=0.75\textwidth]{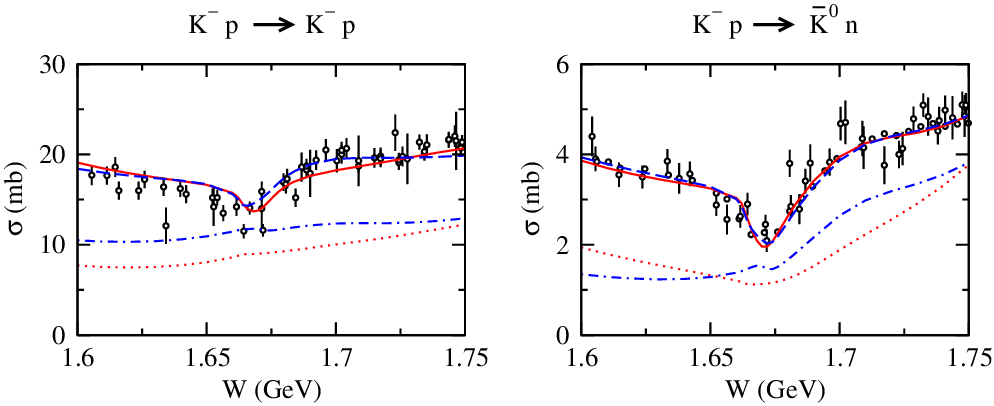}
\caption{(Color online)
The $K^- p \to K^- p$ and $K^- p \to \bar K^0 n$ total cross sections at the energies 
near $W\sim 1.67$ GeV.
Red solid (blue dashed) curves are the full results from Model A (Model B), while
red dotted (blue dashed-dotted) curves are the results for which the $S_{01}$ resonance 
contributions are turned off.
}
\label{fig:Lambda1670-dip}
\end{figure}

\begin{figure}
\includegraphics[clip,width=0.67\textwidth]{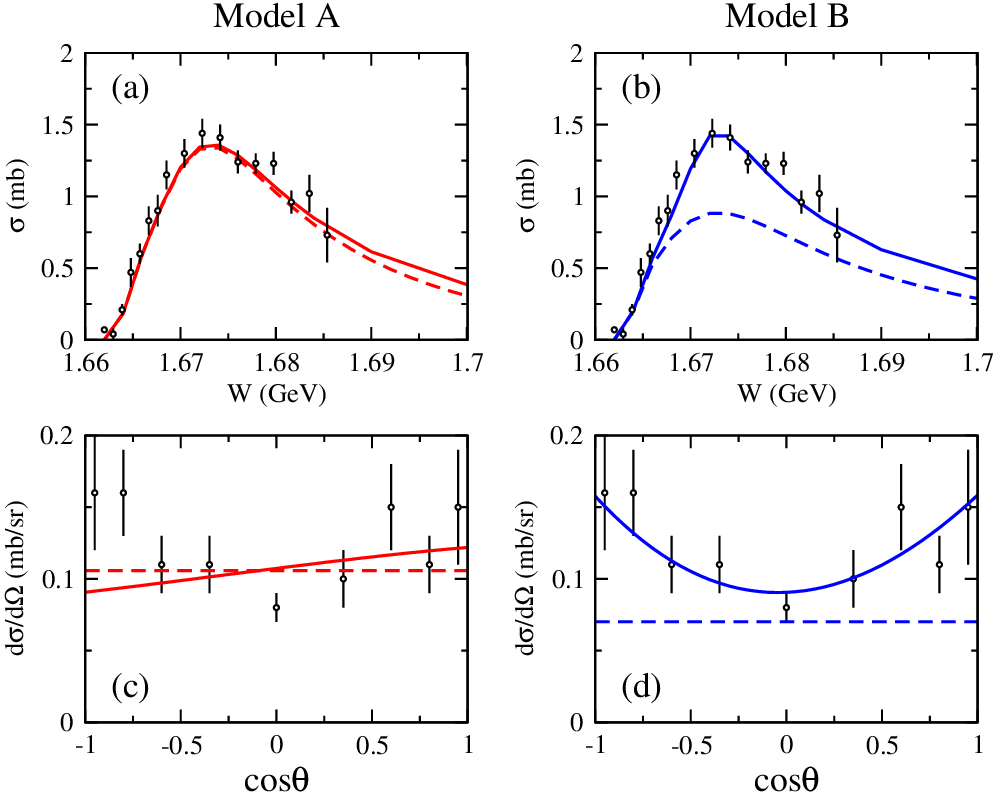}
\caption{(Color online)
Total cross section near the threshold (upper panels) and
differential cross section at $W=1672$ MeV (lower panels) for the $K^- p \to \eta \Lambda$ 
reaction.
Left panels (right panels) are the results from Model A (Model B).
Solid curves are the full results, while the dashed curves are the contribution
from the $S_{01}$ partial wave only.
For Model B, the difference between the solid and dashed curves almost comes from
the $P_{03}$ partial wave dominated by the new narrow 
$J^P=3/2^+$ $\Lambda$ resonance with $M_R = 1671^{+2}_{-8}-i(5^{+11}_{-2})$ MeV.
}
\label{fig:p03-cont}
\end{figure}

In the following, we further discuss each of resonances shown in Figs.~\ref{fig:l-spectrum} 
and~\ref{fig:s-spectrum}:

\begin{description}
\item[$\bm{S_{01}[\Lambda(1/2^-)]}$]
Our two models, Model A and Model B, and the KSU analysis all give a
narrow resonance with $\mathrm{Re}(M_R)\sim 1.67$ GeV (Fig.~\ref{fig:l-spectrum}).
It can be identified with the four-star $\Lambda(1670)1/2^-$ of PDG (narrow black bar in 
the left-most column).
As discussed in our previous paper~\cite{knlskp1}, this resonance is responsible for 
the sharp peak in the $K^- p \to \eta \Lambda$ total cross section near the threshold. 
This strong near-threshold effect is similar to that of $N^*(1535)1/2^-$ resonance on 
the $\pi N\to \eta N$ reaction. 
The $\Lambda(1670)1/2^-$ is also found to be responsible for a dip\footnote{Note 
that a resonance can appear also as a dip in the cross sections,
depending on the interference with background and/or other 
resonance contributions~\cite{taylor}.} in the 
$K^- p \to \bar K N$ total cross sections at $W\sim 1.67$~GeV. This is 
presented in Fig.~\ref{fig:Lambda1670-dip}.
Other than the agreement in extracting the $\Lambda(1670)1/2^-$ resonance
between the three analyses, a broad resonance at $\mathrm{Re}(M_R)\sim 1.5$ GeV
is found in Model B and  two additional resonances 
are found at higher energies in the KSU analysis.
\item[$\bm{P_{01}[\Lambda(1/2^+)]}$]
The lowest resonance at $\mathrm{Re}(M_R)\sim 1.55$ GeV shows an
agreement between the three analyses. 
This resonance would correspond to $\Lambda(1600)1/2^+$, a three-star resonance rated by PDG.
Clearly, the higher resonances are not well determined.
\item[$\bm{P_{03}[\Lambda(3/2^+)]}$]
A resonance  at $\mathrm{Re}(M_R)\sim 1.86$ GeV
is found in Model A and the KSU analysis, but not in Model B.
Thus the current data are not sufficient to establish this state model independently.
If this resonance corresponds to the four-star $\Lambda(1890)3/2^+$ of PDG,
then this is one example indicating that
a four-star resonance rated by PDG using the Breit-Wigner parameters
is not confirmed by the analyses in which the resonance parameters are extracted at pole positions.

The main feature of Model B is 
to  have a  new narrow $\Lambda$ resonance 
with $M_R = 1671^{+2}_{-8}-i(5^{+11}_{-2})$ MeV. It locates in the energy region
close to the $S_{01}$ resonance $\Lambda(1670)1/2^-$ discussed above.
As already discussed in our previous paper~\cite{knlskp1}, the evidence of 
this new narrow resonance could be seen 
from the $K^- p \to \eta \Lambda$ cross sections near the threshold.
To see this, we compare the contributions from this resonance and
the $S_{01}$ resonance $\Lambda(1670)1/2^-$ to
the cross sections. As shown in Fig.~\ref{fig:p03-cont}(a), 
the peak of the $K^- p \to \eta \Lambda$ total cross section 
near the threshold calculated from Model A
is completely dominated by the contribution from the $S_{01}$ partial wave 
that is almost entirely due to the  $\Lambda(1670)1/2^-$ resonance. 
On the other hand,  we see in Fig.~\ref{fig:p03-cont}(b) that
the contribution of the $S_{01}$ partial wave in Model B
 is just about 60 \%, and the remaining 40 \% come almost entirely from 
the $P_{03}$ partial wave that contains this new narrow resonance.
Since both models reproduce the total cross section very well,
the existence of this  new narrow $J^P=3/2^+$ $\Lambda$ resonance cannot be
established only by considering the total cross sections.
To get a deeper insight, it is necessary to at least examine its effects on
the differential cross sections.
The $K^- p \to \eta \Lambda$ differential cross section data near the threshold ($W \sim 1.67$ GeV)
show a clear concave-up angular dependence that cannot be described by the $S$-wave amplitudes.
We see in Fig.~\ref{fig:p03-cont}(c) that the  results (solid curve) from
Model A, which is mainly the $S_{01}$ wave (dashed curve), 
do not reproduce the angular dependence well.
On the other hand,  the new narrow $J^P=3/2^+$ $\Lambda$ resonance
extracted within Model B is found to be responsible for the 
reproduction of the differential cross section 
data.  This is shown in Fig.~\ref{fig:p03-cont}(d), 
suggesting that the angular dependence of the data seems to favor the existence of this new resonance. 
\item[$\bm{D_{03}[\Lambda(3/2^-)]}$]
The first and second resonances in the $D_{03}$ partial wave
extracted from Models A and B and KSU analysis agree very well.
Compared with the PDG values (the left-most column),
the first resonance can be identified with the well known four-star $\Lambda(1520)3/2^-$,
and the second resonance could correspond to the four-star $\Lambda(1690)3/2^-$.
The KSU analysis gives an additional ``new'' resonance with the pole mass $1985 -i223.5$~MeV.
We are not able to confirm this since the imaginary part 
of this resonance pole would correspond to a very large total width
and this resonance is perhaps outside the complex energy 
region considered in our search.
\item[$\bm{D_{05}[\Lambda(5/2^-)]}$]
Model A finds two resonances
for this partial wave, while only one resonance is found in Model B and the KSU analysis.
Although the values of resonance masses from the three analyses are fluctuating,
the resonances found in Model B and the KSU analysis and
the narrower second resonance in Model A might correspond to 
the four-star $\Lambda(1830)5/2^-$ of PDG (left-most column).
\item[$\bm{F_{05}[\Lambda(5/2^+)]}$]
The first resonance extracted from the three analyses agree very well. 
This resonance could correspond to the four-star $\Lambda(1820)5/2^+$ listed by PDG.
We however do not find the broad resonance
with $\mathrm{Re}(M_R)\sim 1.97$ GeV found in the KSU analysis.
\item[$\bm{F_{07}[\Lambda(7/2^+)]}$]
All of the three analyses find one resonance below $\mathrm{Re}(M_R)= 2.1$ GeV. 
However, the real part of its pole mass in Model A
is about 250 MeV lower than that of Model B and the KSU analysis.
Since the resonances found in Models A and B have large uncertainties as shown in 
Table~\ref{tab:res-pole}, at this stage it is difficult to make a conclusion 
for the resonances in this partial wave.
\item[$\bm{S_{11}[\Sigma(1/2^-)]}$]
A sizable analysis dependence of the extracted resonance spectrum is seen in this
partial wave.
A resonance at $\mathrm{Re}(M_R)\sim 1.7$ GeV is found in Model A and the KSU analysis,
which may correspond to $\Sigma(1750)1/2^-$ rated as three-star in PDG.
However, this resonance is not found in Model B.
This may be understood from the fact that 
the energy dependence of the $S_{11}$ partial-wave amplitudes for
$\bar K N \to \bar K N, \pi\Sigma, \pi\Lambda$ 
have rapid changes  at $W\sim 1.7$~GeV
in Model A and the KSU single energy solution, but are rather smooth 
in  Model B (see Figs. 24, 26, and 27 in Ref.~\cite{knlskp1}).
It is interesting to see that Model B and the KSU analysis give a low-lying
resonance with $\mathrm{Re}(M_R) \lesssim 1.6$ GeV.
This might correspond to $\Sigma(1620)1/2^-$ rated as two-star by PDG. 
\item[$\bm{P_{11}[\Sigma(1/2^+)]}$]
Similar to the $S_{11}$ case, the extracted resonance spectrum in
this partial wave also varies sizably between the three analyses.
It is worthwhile to mention that a resonance with a low mass 
$\mathrm{Re}(M_R)< 1.55$ GeV is found in both Model A and Model B.
This might correspond to $\Sigma(1480)$ or $\Sigma(1560)$ in PDG, 
whose evidence is still poor and spin-parity has not been determined 
(and thus not shown in Fig.~\ref{fig:s-spectrum}).
\item[$\bm{P_{13}[\Sigma(3/2^+)]}$]
It is well known that the decuplet $\Sigma(1385)3/2^+$ exists below the $\bar K N$ threshold
in this partial wave. 
To account for the existence of this well-established resonance, 
we set a pole with $M_R=1381-i20$~MeV (not shown in Fig.~\ref{fig:s-spectrum})
in this partial wave as an ``input data'' to constrain parameters of our model Hamiltonian~\cite{knlskp1}. 
The resulting Models A and B, however, do not have any resonances
above the $\bar K N$ threshold.
[Note that the resonance parameters for all of 
the resonances other than the decuplet $\Sigma(1385)3/2^+$
are purely the ``output'' of our reaction models.]
In contrast, the KSU analysis finds two ``new'' resonances
as seen in the panel for ``$3/2^+(P_{13})$'' of Fig.~\ref{fig:s-spectrum}. 
This analysis dependence is perhaps due to the fact that 
the $K^- p \to \bar K N, \pi\Sigma, \pi\Lambda, K\Xi$ reaction data
included in our fits are not sensitive to the $P_{13}$ wave.
Data for $2\to 3$ reactions such as $K^- p \to \pi \pi \Lambda$ and $K^- p \to \pi \bar K N$
may be needed to resolve the analysis dependence.
\item[$\bm{D_{13}[\Sigma(3/2^-)]}$]
All three analyses find a resonance at $\mathrm{Re}(M_R)\sim 1.67$ GeV, 
which would correspond to the four-star $\Sigma(1670) 3/2^-$ of PDG.
It has been suggested that there exists another $\Sigma$ resonance with the same mass 
and quantum numbers as $\Sigma(1670) 3/2^-$ and, in contrast to $\Sigma(1670) 3/2^-$,
this resonance  has a large branching fraction to $\pi\Lambda(1405) \to \pi\pi\Sigma$  
(see the discussions in pp. 1481-1482 of Ref.~\cite{pdg2014} and references therein).
Although the three analyses do not find such an additional resonance,
its existence can be examined conclusively only when
the $\pi\Lambda(1405)$ channel and the data associated with 
the three-body $\pi\pi\Sigma$ production reactions are accounted for in the analysis.
In addition to $\Sigma(1670) 3/2^-$, a resonance 
with lower $\mathrm{Re}(M_R)$ is found in Models A and B.
This resonance may correspond to $\Sigma (1580)3/2^-$ that is rated as one-star in PDG.
\item[$\bm{D_{15}[\Sigma(5/2^-)]}$]
Only one resonance at $\mathrm{Re}(M_R)\sim 1.77$ GeV in this partial wave
is found in Models A and B and the KSU analysis. 
This resonance could correspond to the four-star $\Sigma(1775)5/2^-$ of PDG.
This excellent agreement between the three analyses
strongly suggests that the resonance spectrum of this partial wave
is well established  up to $\mathrm{Re}(M_R)\sim 1.8$ GeV, since
the $D_{15}$ partial-wave amplitudes for $\bar K N \to \bar K N, \pi \Sigma, \pi \Lambda$
are well determined~\cite{knlskp1}.
\item[$\bm{F_{15}[\Sigma(5/2^+)]}$]
All three analyses find one resonance below $\mathrm{Re}(M_R)= 2.1$ GeV.
The real parts of the resonance pole masses are $\mathrm{Re}(M_R) \sim 1.89$ GeV 
for Model A and the KSU analysis, while $\mathrm{Re}(M_R) \sim 1.7$ GeV for Model B,
showing a clear analysis dependence for the extracted pole masses.
If the resonances found in Model A and the KSU analysis
correspond to the four-star $\Sigma(1915)5/2^+$ of PDG,
then this is another example indicating that
a four-star resonance rated by PDG using the Breit-Wigner parameters
is not confirmed by the analyses in which the resonance parameters are extracted at pole positions.
\item[$\bm{F_{17}[\Sigma(7/2^+)]}$]
Only one resonance at $\mathrm{Re}(M_R) \sim 2.02$ GeV is found in
all three analyses.
This resonance could correspond to the four-star $\Sigma(2030)7/2^+$ of PDG.
\end{description}

Before closing this subsection, it is worthwhile to mention that
the $S$-wave resonance poles located near the threshold of 
a two-body channel have a strong correlation with the values of the scattering length and
effective range of the channel, as discussed in Ref.~\cite{effective_range}.
We can examine this by making use of the $S_{01}$ $\Lambda(1670)1/2^-$ resonance that locates close
to the $\eta \Lambda$ threshold.
Near the threshold, the $S$-wave $\eta \Lambda$ scattering amplitudes
can be written as
\begin{equation}
F^{S\mathrm{wave}}_{\eta\Lambda,\eta\Lambda}(k) \simeq k\times
\left(\frac{1}{a_{\eta \Lambda}}-ik+\frac{r_{\eta \Lambda}}{2}k^2 \right)^{-1} ,
\label{eq:eff-exp}
\end{equation}
where the ${\cal O}(k^4)$ terms are neglected in the denominator; and
$a_{\eta \Lambda}$ and $r_{\eta \Lambda}$ are the scattering length and effective range
for the $\eta \Lambda$ scattering, respectively.
These threshold parameters have been extracted in our previous paper~\cite{knlskp1}, 
and their values are:
\begin{equation}
a_{\eta \Lambda} =
\left\{
\begin{array}{cc}
 1.35+i0.36~\mathrm{fm} & (\mathrm{Model~A}),\\
 0.97+i0.51~\mathrm{fm} & (\mathrm{Model~B}),
\end{array}
\right.
\end{equation}
\begin{equation}
r_{\eta \Lambda} =
\left\{
\begin{array}{cc}
-5.67-i2.24~\mathrm{fm} & (\mathrm{Model~A}),\\
-5.82-i3.32~\mathrm{fm} & (\mathrm{Model~B}).
\end{array}
\right.
\end{equation}
Substituting the above values to Eq.~(\ref{eq:eff-exp}),
we find that the approximated scattering amplitude~(\ref{eq:eff-exp})
has a pole in the nearest Riemann energy surface at the on-shell momentum with
$k =73.81-i57.65$ MeV for Model A and with $k =72.03-i72.87$ MeV for Model B.
This means that the amplitude has a pole at the complex $W$ with
\begin{equation}
W = E_{\eta}(k) + E_\Lambda(k) =
\left\{
\begin{array}{cc}
1667-i12~\mathrm{MeV} & (\mathrm{Model~A}),\\
1664-i14~\mathrm{MeV} & (\mathrm{Model~B}).
\end{array}
\right.
\end{equation}
These values indeed show a good agreement with the exact pole values:
$M_R=1669^{+3}_{-8}-i(9^{+9}_{-1})$ MeV for Model A and $M_R=1667^{+1}_{-1}-i(12^{+3}_{-1})$ MeV for Model B.

\subsection{Residues and branching ratios}

Within the Hamiltonian formulation~\cite{msl07} of the dynamical model employed in our analysis,
it can be shown that the residues defined by Eq.~(\ref{eq:F-pole}) can be written as 
$R_{M'B',MB}=\sqrt{\rho_{M'B'}(k_{M'B'}^{\textrm{on}};M_R)} \bra{M'B'}H'\ket{\psi^{R}_{Y^*}}\bra{\psi^{R}_{Y^*}}H'\ket{MB}\sqrt{\rho_{MB}(k_{MB}^{\textrm{on}};M_R)}$,
where $H'$ is the interaction Hamiltonian, and
$\ket{\psi^{R}_{Y^*}}$ is an eigenstate of the total Hamiltonian 
$H\ket{\psi^R_{Y^*}}=M_R\ket{\psi^R_{Y^*}}$
with the outgoing wave boundary condition. 
Since $\bra{\psi^{R}_{Y^*}}H'\ket{MB}$ is related to the strength 
for the transition between a resonance $Y^*$ and a meson-baryon continuum state $MB$,
these resonance parameters contain important information on the structure of the extracted resonances.
The residues $R_{MB,\bar KN}$ extracted from   
the $\bar K N \to MB$ amplitudes within Model~A (Model~B) are
listed in Tables~\ref{tab:a-residue-1} and~\ref{tab:a-residue-2}
(Tables~\ref{tab:b-residue-1} and~\ref{tab:b-residue-2}).
Here, each resonance is specified by its quantum numbers and the value of 
the real part of its pole mass $M_R$.
The residues for the stable channels, 
$MB =\bar K N,\pi\Sigma,\pi\Lambda,\eta\Lambda,K\Xi$, can be evaluated rather straightforwardly 
using the formulas in Sec.~\ref{sec:formulas}.
However, an additional assumption is needed to present our results for the quasi two-body channels,
$MB =\pi\Sigma^*,\bar K^* N$.
These channels  decay into  three-body final states. Thus
their on-shell momenta,  defined by two independent variables,
cannot be determined uniquely at the pole position $W=M_R$.
Strictly speaking, the formulas in Sec.~\ref{sec:formulas} 
cannot be used straightforwardly for the quasi two-body channels.
Therefore, we have taken the following approximate procedure
for $MB =\pi\Sigma^*,\bar K^* N$.
First we recall that 
the dressed $\Sigma^*$ ($\bar K^*$) mass for the $\pi\Sigma^*$ ($\bar K^* N$) channel
within our model is  $1381-i20$ MeV ($899.3-i29.7$ MeV)~\cite{knlskp1}. 
Since the imaginary parts of their masses are small,
we assume that $\Sigma^*$ and $\bar K^*$ appearing in the $\pi\Sigma^*$ and $\bar K^* N$ channels
are ``stable'' particles with the masses 1381 MeV and 899.3 MeV, respectively.
The on-shell momentum for the $\pi\Sigma^*$ and $\bar K^* N$ channels are then
uniquely determined and the residues associated with these two channels
can be computed by using the formulas in Sec.~\ref{sec:formulas}. These results are 
listed in Tables~\ref{tab:a-residue-2} and~\ref{tab:b-residue-2}.
It is noted that a similar approximate procedure was also taken in Ref.~\cite{juelich12} for the
evaluation of the residues associated with $\pi N \to \pi\Delta$.
\begin{table}
\caption{\label{tab:a-residue-1} 
Residues $R_{MB,\bar K N}$ for the stable channels $MB = \bar K N, \pi \Sigma, \pi \Lambda, \eta \Lambda, K\Xi$.
The values presented are of the resonances extracted from Model A.
The magnitude [$R$ (MeV)] and phase [$\phi$ (degree), taken to be $-180^\circ < \phi \leq 180^\circ$]
of $R_{MB,\bar K N} \equiv Re^{i\phi}$ are listed.
Each resonance is specified by the real part of the pole mass $\mathrm{Re}(M_{R})$
and its quantum numbers.
}
\begin{ruledtabular}
\begin{tabular}{lrrrrrrrr}
& 
\multicolumn{2}{c}{$R_{\bar K N,\bar K N}$} &
\multicolumn{2}{c}{$R_{\pi \Sigma,\bar K N}$} &
\multicolumn{2}{c}{$R_{\eta\Lambda,\bar K N}$} &
\multicolumn{2}{c}{$R_{K\Xi,\bar K N}$} \\
\cline{2-3} \cline{4-5} \cline{6-7} \cline{8-9}
Particle $J^P(l_{I2J})$
& $R$ & $\phi$ & $R$ & $\phi$ & $R$ & $\phi$ & $R$ & $\phi$ \\
\hline
$\Lambda(1669)1/2^-(S_{01})$&   3.33&   164&   3.10&   125&   4.49&    59&      -&     -\\
$\Lambda(1544)1/2^+(P_{01})$&   5.86& $-$80&  12.98&   108&      -&     -&      -&     -\\
$\Lambda(2097)1/2^+(P_{01})$&  17.05& $-$63&   2.70&    29&  12.91&   165&   7.80& $-$64\\
$\Lambda(1859)3/2^+(P_{03})$&  13.62& $-$23&   5.70&   104&   2.74& $-$54&   3.17& $-$85\\
$\Lambda(1517)3/2^-(D_{03})$&   3.29& $-$11&   3.32& $-$10&      -&     -&      -&     -\\
$\Lambda(1697)3/2^-(D_{03})$&   8.19&     3&  10.28&$-$173&   0.19&    81&      -&     -\\
$\Lambda(1766)5/2^-(D_{05})$&   1.50&$-$116&  10.42&   102&   1.27&    91&      -&     -\\
$\Lambda(1899)5/2^-(D_{05})$&   0.20& $-$80&   0.23&   179&   0.38& $-$65&   1.91&    94\\
$\Lambda(1824)5/2^+(F_{05})$&  21.48& $-$13&  13.74&   168&   0.71&  $-$3&   0.04&    70\\
$\Lambda(1757)7/2^+(F_{07})$&   0.01& $-$77&   0.81&   120&   0.06&$-$100&      -&     -\\
\\
& 
\multicolumn{2}{c}{$R_{\bar K N,\bar K N}$} &
\multicolumn{2}{c}{$R_{\pi \Sigma,\bar K N}$} &
\multicolumn{2}{c}{$R_{\pi\Lambda,\bar K N}$} &
\multicolumn{2}{c}{$R_{K\Xi,\bar K N}$} \\
\cline{2-3} \cline{4-5} \cline{6-7} \cline{8-9}
Particle $J^P(l_{I2J})$
& $R$ & $\phi$ & $R$ & $\phi$ & $R$ & $\phi$ & $R$ & $\phi$ \\
\hline
$\Sigma(1704)1/2^-(S_{11})$&   4.25&   178&   8.32&   137&   8.93&   169&      -&     -\\
$\Sigma(1547)1/2^+(P_{11})$&   2.27&   168&  14.68&    78&   5.63& $-$84&      -&     -\\
$\Sigma(1706)1/2^+(P_{11})$&   1.35&    91&   7.35&$-$171&   5.90& $-$76&      -&     -\\
$\Sigma(1607)3/2^-(D_{13})$&   0.98&    51&   7.90&  $-$6&   7.46&   156&      -&     -\\
$\Sigma(1669)3/2^-(D_{13})$&   4.13& $-$20&   7.97& $-$21&   2.61&  $-$7&      -&     -\\
$\Sigma(1767)5/2^-(D_{15})$&  23.78& $-$32&   7.36& $-$24&  20.85&   157&      -&     -\\
$\Sigma(1890)5/2^+(F_{15})$&   1.90& $-$15&   7.64&   157&   3.67&   166&   0.10& $-$88\\
$\Sigma(2025)7/2^+(F_{17})$&  14.32& $-$38&   5.24&   135&   8.96& $-$24&   2.26&   129\\
\end{tabular}
\end{ruledtabular}
\end{table}
\begin{table}
\caption{\label{tab:a-residue-2} 
Residues $R_{MB,\bar K N}$ for the unstable channels $MB= \pi\Sigma^*, \bar K^* N$. 
The values presented are of the resonances extracted from Model A.
The magnitude [$R$ (MeV)] and phase [$\phi$ (degree), taken to be $-180^\circ < \phi \leq 180^\circ$]
of $R_{MB,\bar K N} \equiv Re^{i\phi}$ are listed.
Each resonance is specified by the real part of the pole mass $\mathrm{Re}(M_{R})$
and its quantum numbers.
The quantum numbers for the $(\pi\Sigma^*)_i$ ($i=1,2$) and $(\bar K^* N)_i$ ($i=1,2,3$)
channels for a given $J^P$ are presented in Table~\ref{tab:qn}
}
\begin{ruledtabular}
\begin{tabular}{lrrrrrrrrrr}
& 
\multicolumn{2}{c}{$R_{(\pi\Sigma^*)_1,\bar K N}$} &
\multicolumn{2}{c}{$R_{(\pi\Sigma^*)_2,\bar K N}$} &
\multicolumn{2}{c}{$R_{(\bar K^* N)_1,\bar K N}$} &
\multicolumn{2}{c}{$R_{(\bar K^* N)_2,\bar K N}$} &
\multicolumn{2}{c}{$R_{(\bar K^* N)_3,\bar K N}$} \\
\cline{2-3} \cline{4-5} \cline{6-7} \cline{8-9} \cline{10-11}
Particle $J^P(l_{I2J})$
& $R$ & $\phi$ & $R$ & $\phi$ & $R$ & $\phi$ & $R$ & $\phi$ & $R$ & $\phi$ \\
\hline
$\Lambda(1669)1/2^-(S_{01})$&   0.94&$-$104&      -&     -&      -&     -&      -&     -&     -&     -\\
$\Lambda(1544)1/2^+(P_{01})$&  10.21&    77&      -&     -&      -&     -&      -&     -&     -&     -\\
$\Lambda(2097)1/2^+(P_{01})$&  20.32&$-$103&      -&     -&  13.24& $-$97&   4.14&     2&     -&     -\\
$\Lambda(1859)3/2^+(P_{03})$&  16.65& $-$40&   3.61&   127&  10.63&$-$160&  11.80&    15&  0.79&   129\\
$\Lambda(1517)3/2^-(D_{03})$&   3.29&$-$123&   0.11&   122&      -&     -&      -&     -&     -&     -\\
$\Lambda(1697)3/2^-(D_{03})$&   4.37&   168&  10.42& $-$22&      -&     -&      -&     -&     -&     -\\
$\Lambda(1766)5/2^-(D_{05})$&   8.50&    87&   0.43&$-$109&      -&     -&      -&     -&     -&     -\\
$\Lambda(1899)5/2^-(D_{05})$&   0.95&   113&   0.03&   127&   1.11&$-$177&   1.02&     3&  0.31& $-$17\\
$\Lambda(1824)5/2^+(F_{05})$&  13.11&   161&   7.75&   151&   0.29&    41&   6.58&$-$139&  0.02&   161\\
$\Lambda(1757)7/2^+(F_{07})$&   0.33& $-$82&  0.002&$-$128&      -&     -&      -&     -&     -&     -\\
\\
& 
\multicolumn{2}{c}{$R_{(\pi\Sigma^*)_1,\bar K N}$} &
\multicolumn{2}{c}{$R_{(\pi\Sigma^*)_2,\bar K N}$} &
\multicolumn{2}{c}{$R_{(\bar K^* N)_1,\bar K N}$} &
\multicolumn{2}{c}{$R_{(\bar K^* N)_2,\bar K N}$} &
\multicolumn{2}{c}{$R_{(\bar K^* N)_3,\bar K N}$} \\
\cline{2-3} \cline{4-5} \cline{6-7} \cline{8-9} \cline{10-11}
Particle $J^P(l_{I2J})$
& $R$ & $\phi$ & $R$ & $\phi$ & $R$ & $\phi$ & $R$ & $\phi$ & $R$ & $\phi$ \\
\hline
$\Sigma(1704)1/2^-(S_{11})$&   2.31&    73&      -&     -&      -&     -&      -&     -&     -&     -\\
$\Sigma(1547)1/2^+(P_{11})$&   4.71& $-$44&      -&     -&      -&     -&      -&     -&     -&     -\\
$\Sigma(1706)1/2^+(P_{11})$&   3.65&$-$128&      -&     -&      -&     -&      -&     -&     -&     -\\
$\Sigma(1607)3/2^-(D_{13})$&   4.65& $-$18&   1.31&   123&      -&     -&      -&     -&     -&     -\\
$\Sigma(1669)3/2^-(D_{13})$&   7.30&   167&   2.93&   141&      -&     -&      -&     -&     -&     -\\
$\Sigma(1767)5/2^-(D_{15})$&  25.05&   137&   0.83& $-$58&      -&     -&      -&     -&     -&     -\\
$\Sigma(1890)5/2^+(F_{15})$&   3.51&   161&   0.79&$-$163&   0.23&     4&   2.40&    51&  0.02&    16\\
$\Sigma(2025)7/2^+(F_{17})$&   5.78& $-$23&   1.59&   132&  12.54&    38&  20.76&    37&  0.23&    22\\
\end{tabular}
\end{ruledtabular}
\end{table}
\begin{table}
\caption{\label{tab:b-residue-1} 
Residues $R_{MB,\bar K N}$ for the stable channels $MB = \bar K N, \pi \Sigma, \pi \Lambda, \eta \Lambda, K\Xi$.
The values presented are of the resonances extracted from Model B.
See the caption of Table~\ref{tab:a-residue-1} for the description of the table.
}
\begin{ruledtabular}
\begin{tabular}{lrrrrrrrr}
& 
\multicolumn{2}{c}{$R_{\bar K N,\bar K N}$} &
\multicolumn{2}{c}{$R_{\pi \Sigma,\bar K N}$} &
\multicolumn{2}{c}{$R_{\eta \Lambda,\bar K N}$} &
\multicolumn{2}{c}{$R_{K\Xi,\bar K N}$} \\
\cline{2-3} \cline{4-5} \cline{6-7} \cline{8-9}
Particle $J^P(l_{I2J})$ 
& $R$ & $\phi$ & $R$ & $\phi$ & $R$ & $\phi$ & $R$ & $\phi$ \\
\hline
$\Lambda(1512)1/2^-(S_{01})$&  21.11&$-$146&  32.36&    44&      -&     -&      -&     -\\
$\Lambda(1667)1/2^-(S_{01})$&   3.26&   160&   3.30&   131&   4.40&    53&      -&     -\\
$\Lambda(1548)1/2^+(P_{01})$&   9.58&$-$120&  21.82&   101&      -&     -&      -&     -\\
$\Lambda(1841)1/2^+(P_{01})$&   3.90& $-$64&   2.43& $-$24&   1.64&    92&   3.62& $-$82\\
$\Lambda(1671)3/2^+(P_{03})$&   0.17&    57&   0.37&    16&   0.61&   172&      -&     -\\
$\Lambda(1517)3/2^-(D_{03})$&   4.06& $-$10&   3.87&  $-$9&      -&     -&      -&     -\\
$\Lambda(1697)3/2^-(D_{03})$&  12.56&  $-$3&  11.45&$-$177&   0.82& $-$47&      -&     -\\
$\Lambda(1924)5/2^-(D_{05})$&   1.78& $-$77&   0.43& $-$75&   0.31& $-$53&   0.59&    69\\
$\Lambda(1821)5/2^+(F_{05})$&  18.74& $-$21&   9.43&   162&   1.84& $-$23&   0.03&   163\\
$\Lambda(2041)7/2^+(F_{07})$&   1.30& $-$51&   7.76& $-$49&   1.34& $-$69&   6.34& $-$79\\
\\
& 
\multicolumn{2}{c}{$R_{\bar K N,\bar K N}$} &
\multicolumn{2}{c}{$R_{\pi \Sigma,\bar K N}$} &
\multicolumn{2}{c}{$R_{\pi \Lambda,\bar K N}$} &
\multicolumn{2}{c}{$R_{K\Xi,\bar K N}$} \\
\cline{2-3} \cline{4-5} \cline{6-7} \cline{8-9}
Particle $J^P(l_{I2J})$ 
& $R$ & $\phi$ & $R$ & $\phi$ & $R$ & $\phi$ & $R$ & $\phi$ \\
\hline
$\Sigma(1551)1/2^-(S_{11})$&  45.58&   131&  23.59&     7&  48.08& $-$38&      -&     -\\
$\Sigma(1940)1/2^-(S_{11})$&  43.48&    57&  22.00&    54&   7.29&    29&   8.61& $-$93\\
$\Sigma(1457)1/2^+(P_{11})$&   1.65& $-$45&   1.30&   172&   8.19&   137&      -&     -\\
$\Sigma(1605)1/2^+(P_{11})$&   8.62& $-$43&  14.35&   131&  17.43&    81&      -&     -\\
$\Sigma(2014)1/2^+(P_{11})$&   9.07&    72&   6.57&    84&   7.75&   144&   4.72&  $-$6\\
$\Sigma(1492)3/2^-(D_{13})$&   0.02&$-$162&   0.34&    56&   0.97&$-$121&      -&     -\\
$\Sigma(1672)3/2^-(D_{13})$&   1.86& $-$20&   7.16&  $-$6&   2.35& $-$37&      -&     -\\
$\Sigma(1765)5/2^-(D_{15})$&  22.61& $-$35&   7.58& $-$36&  17.60&   150&      -&     -\\
$\Sigma(1695)5/2^+(F_{15})$&   0.40& $-$61&   3.91&   110&   3.99&   111&      -&     -\\
$\Sigma(2014)7/2^+(F_{17})$&  22.78& $-$43&   1.27&    45&  14.23& $-$42&   4.41&   116  
\end{tabular}
\end{ruledtabular}
\end{table}
\begin{table}
\caption{\label{tab:b-residue-2} 
Residues $R_{MB,\bar K N}$ for the unstable channels $MB= \pi\Sigma^*, \bar K^* N$. 
The values presented are of the resonances extracted from Model B.
See the caption of Table~\ref{tab:a-residue-2} for the description of the table.
}
\begin{ruledtabular}
\begin{tabular}{lrrrrrrrrrr}
& 
\multicolumn{2}{c}{$R_{(\pi\Sigma^*)_1,\bar K N}$} &
\multicolumn{2}{c}{$R_{(\pi\Sigma^*)_2,\bar K N}$} &
\multicolumn{2}{c}{$R_{(\bar K^* N)_1,\bar K N}$} &
\multicolumn{2}{c}{$R_{(\bar K^* N)_2,\bar K N}$} &
\multicolumn{2}{c}{$R_{(\bar K^* N)_3,\bar K N}$} \\
\cline{2-3} \cline{4-5} \cline{6-7} \cline{8-9} \cline{10-11}
Particle $J^P(L_{I2J})$
& $R$ & $\phi$ & $R$ & $\phi$ & $R$ & $\phi$ & $R$ & $\phi$ & $R$ & $\phi$ \\
\hline
$\Lambda(1512)1/2^-(S_{01})$&   3.52&    16&      -&     -&      -&     -&      -&     -&     -&     -\\
$\Lambda(1667)1/2^-(S_{01})$&   3.16&    74&      -&     -&      -&     -&      -&     -&     -&     -\\
$\Lambda(1548)1/2^+(P_{01})$&  16.39&    51&      -&     -&      -&     -&      -&     -&     -&     -\\
$\Lambda(1841)1/2^+(P_{01})$&   2.27&     2&      -&     -&   1.05& $-$31&   8.73&  $-$5&     -&     -\\
$\Lambda(1671)3/2^+(P_{03})$&   0.55&    14&   0.03&$-$168&      -&     -&      -&     -&     -&     -\\
$\Lambda(1517)3/2^-(D_{03})$&   3.34&$-$123&   0.18&   125&      -&     -&      -&     -&     -&     -\\
$\Lambda(1697)3/2^-(D_{03})$&   4.01&   179&  14.53& $-$26&      -&     -&      -&     -&     -&     -\\
$\Lambda(1924)5/2^-(D_{05})$&   9.00&   125&   0.32& $-$61&   1.61& $-$26&   1.23&   166&  1.55&     6\\
$\Lambda(1821)5/2^+(F_{05})$&  13.23& $-$24&   2.51&   144&   0.28&  $-$1&   6.13&   111&  0.01&$-$175\\
$\Lambda(2041)7/2^+(F_{07})$&   3.59&   103&   0.27&   112&   1.66&     9&   1.86&$-$165&  1.42&   168\\
\\
& 
\multicolumn{2}{c}{$R_{(\pi\Sigma^*)_1,\bar K N}$} &
\multicolumn{2}{c}{$R_{(\pi\Sigma^*)_2,\bar K N}$} &
\multicolumn{2}{c}{$R_{(\bar K^* N)_1,\bar K N}$} &
\multicolumn{2}{c}{$R_{(\bar K^* N)_2,\bar K N}$} &
\multicolumn{2}{c}{$R_{(\bar K^* N)_3,\bar K N}$} \\
\cline{2-3} \cline{4-5} \cline{6-7} \cline{8-9} \cline{10-11}
Particle $J^P(L_{I2J})$
& $R$ & $\phi$ & $R$ & $\phi$ & $R$ & $\phi$ & $R$ & $\phi$ & $R$ & $\phi$ \\
\hline
$\Sigma(1551)1/2^-(S_{11})$&   3.80&$-$176&      -&     -&      -&     -&      -&     -&     -&     -\\
$\Sigma(1940)1/2^-(S_{11})$&  12.47&$-$163&      -&     -&  14.93&   164&  32.21&    82&     -&     -\\
$\Sigma(1457)1/2^+(P_{11})$&      -&     -&      -&     -&      -&     -&      -&     -&     -&     -\\
$\Sigma(1605)1/2^+(P_{11})$&  14.84&    87&      -&     -&      -&     -&      -&     -&     -&     -\\
$\Sigma(2014)1/2^+(P_{11})$&   7.77&    63&      -&     -&  10.70&   137&  14.83& $-$50&     -&     -\\
$\Sigma(1492)3/2^-(D_{13})$&   0.23& $-$54&   0.01& $-$16&      -&     -&      -&     -&     -&     -\\
$\Sigma(1672)3/2^-(D_{13})$&   0.93&    99&   1.32&   125&      -&     -&      -&     -&     -&     -\\ 
$\Sigma(1765)5/2^-(D_{15})$&  25.51&   131&   0.42& $-$58&      -&     -&      -&     -&     -&     -\\
$\Sigma(1695)5/2^+(F_{15})$&   0.39&    97&   0.06&    88&      -&     -&      -&     -&     -&     -\\
$\Sigma(2014)7/2^+(F_{17})$&  37.94& $-$51&   3.68&   114&   5.38&    22&   7.50&    18&  8.05&  $-$9
\end{tabular}
\end{ruledtabular}
\end{table}

\begin{figure}
\includegraphics[clip,width=0.75\textwidth]{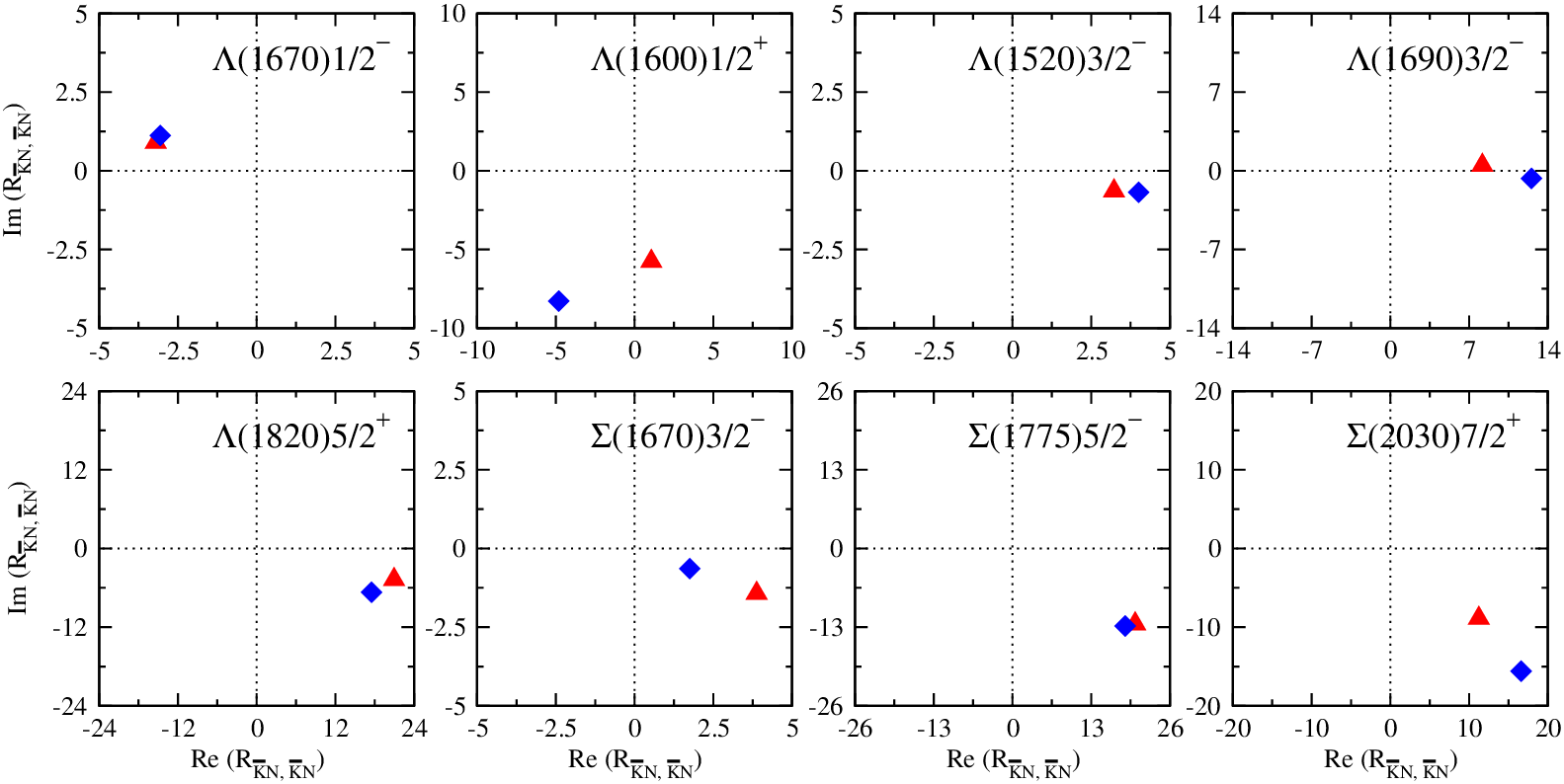}
\caption{(Color online)
Graphical comparison of $R_{\bar K N, \bar KN}$ between Model A and Model B
for the well-established resonances (see text for the details).
}
\label{fig:residue-comp} 
\end{figure}
We see  in Figs.~\ref{fig:l-spectrum} and~\ref{fig:s-spectrum} that
the pole masses of eight resonances, represented in the PDG notation by
$\Lambda(1670)1/2^-$, $\Lambda(1600)1/2^+$, $\Lambda(1520)3/2^-$,
$\Lambda(1690)3/2^-$, $\Lambda(1820)5/2^+$,
$\Sigma(1670)3/2^-$, $\Sigma(1775)5/2^-$, and $\Sigma(2030)7/2^+$,
agree very well between our two models and the KSU analysis.
The residues $R_{\bar KN, \bar KN}$ of these resonances extracted from  Model A and Model B
are compared\footnote{ The KSU analysis did not provide their residues in Ref.~\cite{zhang2013b}.} 
in Fig.~\ref{fig:residue-comp}.
They in general  agree well,
while visible differences are seen for several resonances. 
In particular, the pole mass for $\Sigma(1670)3/2^-$ agrees 
within 1 \% accuracy between Model A and Model B, but
their residues differ by a factor of about 2  in  magnitude.
This implies that the residues are more sensitive to the analysis 
than the pole masses.

We now turn to discussing the branching ratios of the decays of the extracted 
resonances because it may provide us with an intuitive understanding for the properties 
of the resonances.
The branching ratio for the $Y^* \to MB$ decay may be defined as 
$B_{MB}=2|R_{MB,MB}|/\Gamma^{\textrm{tot}}$,
where $R_{MB,MB}$ is the residue for the $MB$ scattering amplitude evaluated at 
the pole position of the considered $Y^*$ resonance and
$\Gamma^{\textrm{tot}} = -2\mathrm{Im}(M_R)$ is the total width of the resonance.
However, it is known that the sum of the branching ratios defined in this manner
do not necessarily equal to unity~\cite{knls13,juelich12}.
This would be somewhat problematic as a notion of ``ratio,'' and 
will require further studies to give a reasonable interpretation to this definition.
Instead, here we will follow the procedures developed in Ref.~\cite{kpi2pi13}
to present the branching ratios
evaluated using the following equations,
\begin{equation}
B_{MB} = \frac{\gamma_{MB}}{\sum_{MB}\gamma_{MB}}.
\label{eq:br-bw}
\end{equation}
Here, the ``partial decay width'' $\gamma_{MB}$ is defined for the stable meson-baryon channels 
($MB=\bar K N, \pi \Sigma, \pi\Lambda, \eta \Lambda, K\Xi$) as
\begin{equation}
\gamma_{MB} = \rho_{MB}(\bar k;\bar M) \left| \bar\Gamma^R_{MB}(\bar k; \bar M)\right|^2,
\label{eq:gamma-stable}
\end{equation}
where $\bar M = \mathrm{Re}(M_R)$,
$\bar k$ is given by $\bar M = E_M(\bar k) + E_B(\bar k)$.
For the quasi two-body channels $\pi\Sigma^*$ and~$\bar K^* N$
that decay into  $\pi\pi\Lambda$ and $\pi\bar K N$, 
respectively, the $\gamma_{MB}$ are given by
\begin{equation}
\gamma_{\pi\Sigma^*} = 
\frac{1}{2\pi}\int^{\bar M-m_\pi}_{m_\pi + m_\Lambda} dM_{\pi \Lambda}
\frac{-2\mathrm{Im}\bm{(}\Sigma_{\pi\Sigma^*}(\bar k;\bar M)\bm{)}}
     {\left|\bar M - E_\pi(\bar k) - E_{\Sigma^*}(\bar k)-\Sigma_{\pi\Sigma^*}(\bar k;\bar M)\right|^2}
 \rho_{\pi\Sigma^*}(\bar k;\bar M) \left| \bar\Gamma^R_{\pi\Sigma^*}(\bar k; \bar M)\right|^2,
\label{eq:gamma-quasi-piss}
\end{equation}
for the case of $MB = \pi \Sigma^*$, and
\begin{equation}
\gamma_{\bar K^* N} = 
\frac{1}{2\pi}\int^{\bar M-m_N}_{m_\pi + m_{\bar K}} dM_{\pi \bar K}
\frac{-2\mathrm{Im}\bm{(}\Sigma_{\bar K^*N}(\bar k;\bar M)\bm{)}}
     {\left|\bar M - E_{\bar K^*}(\bar k) - E_N(\bar k)-\Sigma_{\bar K^*N}(\bar k;\bar M)\right|^2}
 \rho_{\bar K^* N}(\bar k;\bar M) \left| \bar\Gamma^R_{\bar K^* N}(\bar k; \bar M)\right|^2,
\label{eq:gamma-quasi-kbsn}
\end{equation}
for the case of $MB = \bar K^* N$.
Here $\Sigma_{MB}(k;W)$ is the self-energy for the $MB$ Green's function given in Ref.~\cite{knlskp1}; 
$\bar k$ is defined by $\bar M = E_\pi(\bar k) + \sqrt{M_{\pi \Lambda}^2 + \bar k^2}$ 
[$\bar M = E_N(\bar k) + \sqrt{M_{\pi \bar K}^2 + \bar k^2}$]
for $MB = \pi\Sigma^*$ [$MB = \bar K^* N$]; and
the undressed values listed in Table V of Ref.~\cite{knlskp1} are used for the $\Sigma^*$ and $\bar K^*$
masses.
The integrals in Eqs.~(\ref{eq:gamma-quasi-piss}) and~(\ref{eq:gamma-quasi-kbsn})
account for the phase space of the final three-body states. As expected, 
Eqs.~(\ref{eq:gamma-quasi-piss}) and~(\ref{eq:gamma-quasi-kbsn}) are
reduced to Eq.~(\ref{eq:gamma-stable}) for the stable two-body channels
in the limit of 
$\Sigma_{\pi\Sigma^*}\to 0$ and $\Sigma_{\bar K^* N}\to 0$, respectively.

Summing up the branching ratios defined by Eqs.~(\ref{eq:br-bw})-(\ref{eq:gamma-quasi-kbsn}) 
trivially results in unity.  
We have confirmed that the branching ratios defined by
Eqs.~(\ref{eq:br-bw})-(\ref{eq:gamma-quasi-kbsn})
are in good agreement  with the ones defined by $B_{MB}=2|R_{MB,MB}|/\Gamma^{\textrm{tot}}$
if the sum of the latter ratios is within the range between 0.9 and 1.1.

\begin{figure}
\includegraphics[clip,width=0.67\textwidth]{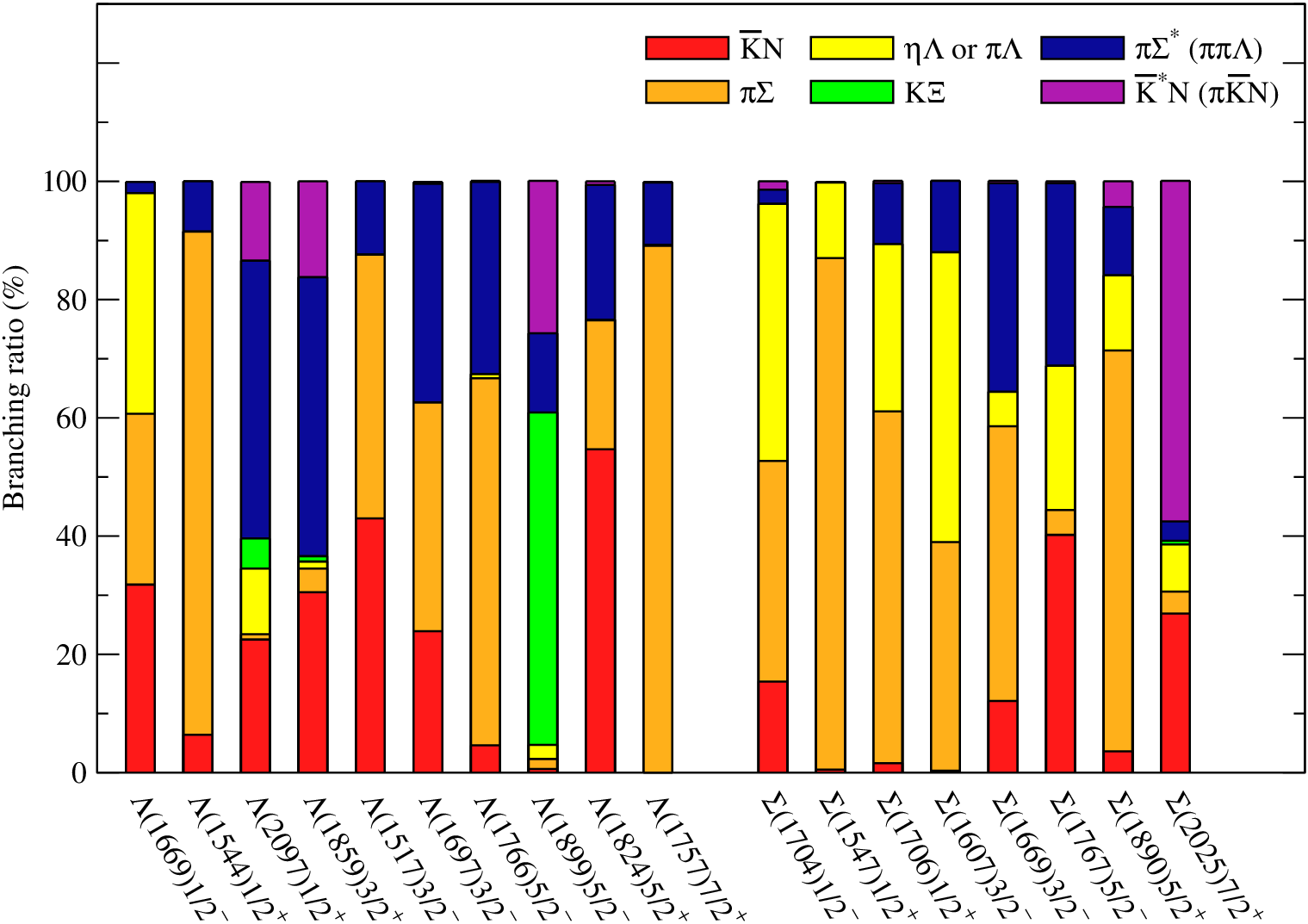}
\caption{(Color online)
Graphical representation of the branching ratios for decays of 
$\Lambda^*$ and $\Sigma^*$ resonances found from Model A.
Equations~(\ref{eq:br-bw})-(\ref{eq:gamma-quasi-kbsn}) are used for evaluating the ratios.
}
\label{fig:br-a}
\end{figure}

\begin{figure}
\includegraphics[clip,width=0.67\textwidth]{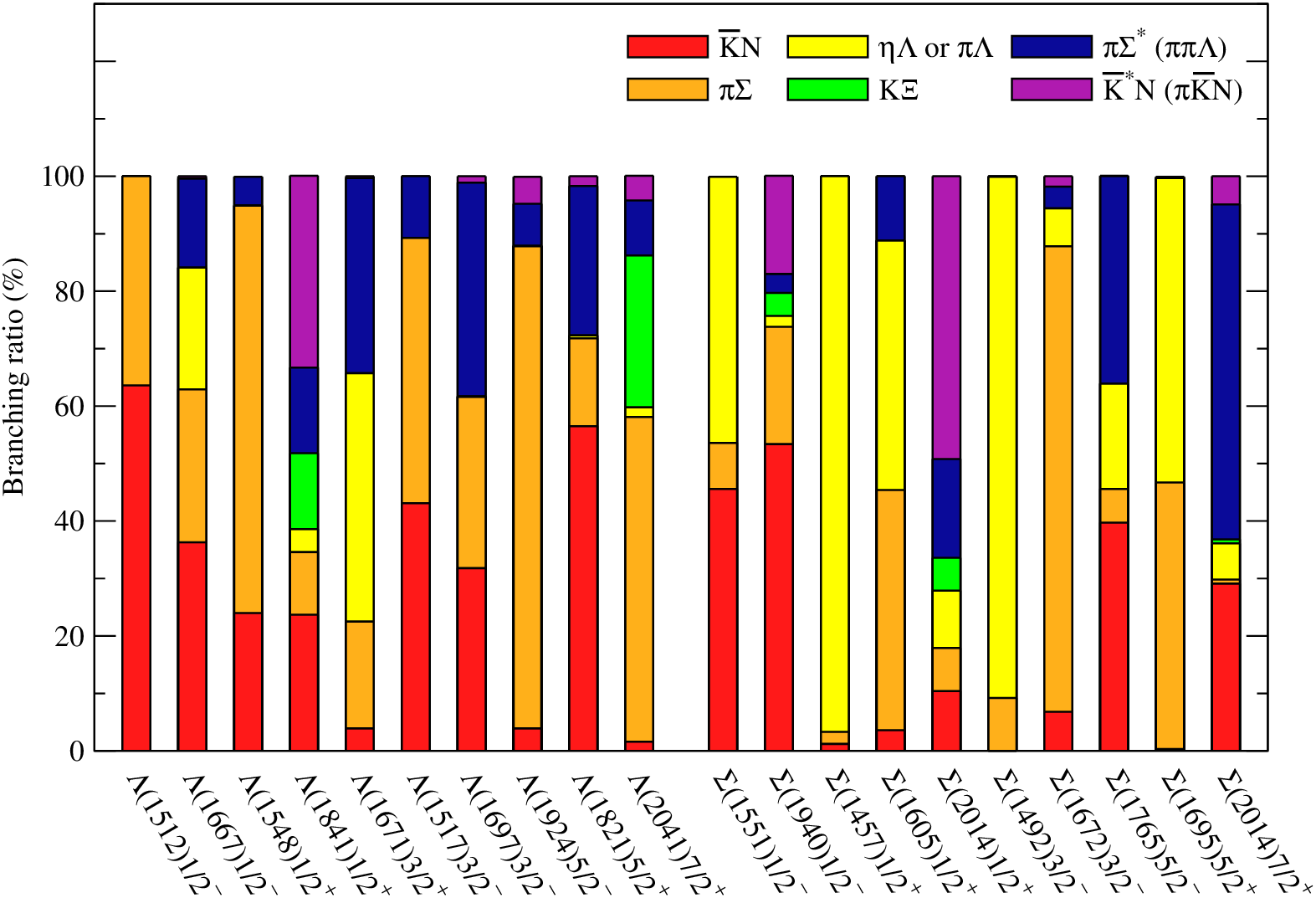}
\caption{(Color online)
Graphical representation of the branching ratios for decays of 
$\Lambda^*$ and $\Sigma^*$ resonances found from Model B.
Equations~(\ref{eq:br-bw})-(\ref{eq:gamma-quasi-kbsn}) are used for evaluating the ratios.
}
\label{fig:br-b}
\end{figure}

The resulting  branching ratios and their graphical representations
are presented in Table~\ref{tab:br-a} and Fig.~\ref{fig:br-a} 
(Table~\ref{tab:br-b} and Fig.~\ref{fig:br-b}) for Model A (Model B).
Except few cases,
the low-mass resonances generally have  large branching ratios of their decays into 
the $\bar K N$ and $\pi \Sigma$ channels.
We also note that the branching ratios to the $\eta\Lambda$ channel of
the narrow $S$-wave $\Lambda$ resonances at $\mathrm{Re}(M_R)\sim 1.67$~GeV,
namely $\Lambda(1669)1/2^-$ for Model A and $\Lambda(1667)1/2^-$ for Model B,
are large and  comparable.
The new narrow $P_{03}$ resonances, $\Lambda(1671)3/2^+$,
found in Model B also has the same feature.
This can be understood from their sizable contributions to the $K^-p \to \eta \Lambda$
total cross sections near the threshold.
Also, the low-lying $\Sigma^*$ resonances found in Model B, 
$\Sigma(1457)1/2^+$ and $\Sigma(1492)3/2^-$, largely decay into the $\pi\Lambda$ channel.
On the other hand, the high-mass resonances are found to have  large branching ratios
to $\pi\Sigma^*$ and $\bar K^* N$ channels, which decay subsequently to the three-body 
$\pi \pi \Lambda$ and $\pi \bar K N$ channels, respectively.
For example, the $J^P=7/2^+$ $\Sigma$ resonance
that would correspond to the four-star $\Sigma(2030)7/2^+$ of PDG, 
namely $\Sigma(2025)7/2^+$ for Model A and $\Sigma(2014)7/2^+$ for Model B,
has a large breaching ratio to the three-body decay channels.
Interestingly, the $J^P=7/2^+$ $\Sigma$ resonance
of Model A mainly decays into $\pi \bar K N$, 
while that of Model B decays into $\pi\pi \Lambda$, 
revealing that our knowledge of the
properties of this four-star resonance is still poor.
This implies a particular importance of the data associated
with three-body channels for establishing the high-mass $\Lambda^*$ and $\Sigma^*$ 
spectrum and their internal structures.
This is quite similar to the case of the $N^*$ and $\Delta^*$ spectroscopy, 
where the data associated with the three-body $\pi \pi N$ channel are expected to play a crucial role 
for establishing the high-mass $N^*$ and $\Delta^*$ resonance mass spectrum, see, e.g., 
Ref.~\cite{kpi2pi13}.
\begin{table}
\caption{\label{tab:br-a} 
Branching ratios for the decays of $\Lambda^*$ and $\Sigma^*$ resonances extracted from Model A.
Equations~(\ref{eq:br-bw})-(\ref{eq:gamma-quasi-kbsn}) are used for evaluating the ratios.
The quantum numbers for the $(\pi\Sigma^*)_i$ ($i=1,2$) and $(\bar K^* N)_i$ ($i=1,2,3$)
channels for a given $J^P$ are presented in Table~\ref{tab:qn}
}
{\footnotesize
\begin{ruledtabular}
\begin{tabular}{lrrrrrrrrr}
&\multicolumn{9}{c}{Branching ratios (\%)}\\
\cline{2-10}
Particle $J^P(l_{I2J})$ & 
$B_{\bar K N}$&
$B_{\pi \Sigma}$&
$B_{\eta \Lambda}$&
$B_{K \Xi}$&
$B_{(\pi\Sigma^*)_1}$&
$B_{(\pi\Sigma^*)_2}$&
$B_{(\bar K^* N)_1}$&
$B_{(\bar K^* N)_2}$&
$B_{(\bar K^* N)_3}$\\
\hline
$\Lambda(1669)1/2^-(S_{01})$& 31.8& 28.9& 37.3&    -&  1.9&    -&  0.0&  0.0&    -\\
$\Lambda(1544)1/2^+(P_{01})$&  6.4& 85.1&    -&    -&  8.5&    -&    -&    -&    -\\
$\Lambda(2097)1/2^+(P_{01})$& 22.5&  0.9& 11.1&  5.1& 47.0& 13.0&  0.3&    -&    -\\
$\Lambda(1859)3/2^+(P_{03})$& 30.5&  4.0&  1.2&  0.9& 45.3&  1.9&  7.3&  8.8&  0.1\\
$\Lambda(1517)3/2^-(D_{03})$& 43.0& 44.6&    -&    -& 12.1&  0.3&    -&    -&    -\\
$\Lambda(1697)3/2^-(D_{03})$& 23.9& 38.7&  0.0&    -&  6.2& 30.8&  0.0&  0.3&  0.0\\
$\Lambda(1766)5/2^-(D_{05})$&  4.6& 62.1&  0.7&    -& 32.4&  0.1&  0.1&  0.1&  0.0\\
$\Lambda(1899)5/2^-(D_{05})$&  0.6&  1.7&  2.4& 56.2& 13.4&  0.0& 13.4& 11.5&  0.9\\
$\Lambda(1824)5/2^+(F_{05})$& 54.7& 21.8&  0.1&  0.0& 17.3&  5.5&  0.0&  0.6&  0.0\\
$\Lambda(1757)7/2^+(F_{07})$&  0.0& 89.1&  0.2&    -& 10.5&  0.0&  0.0&  0.1&  0.0\\
\\                                                                        
Particle $J^P(l_{I2J})$ &
$B_{\bar K N}$&
$B_{\pi \Sigma}$&
$B_{\pi \Lambda}$&
$B_{K \Xi}$&
$B_{(\pi\Sigma^*)_1}$&
$B_{(\pi\Sigma^*)_2}$&
$B_{(\bar K^* N)_1}$&
$B_{(\bar K^* N)_2}$&
$B_{(\bar K^* N)_3}$\\
\hline
$\Sigma(1704)1/2^-(S_{11})$& 15.4& 37.3& 43.5&    -&  2.4&    -&  0.4&  1.0&    -\\
$\Sigma(1547)1/2^+(P_{11})$&  0.5& 86.5& 12.8&    -&  0.1&    -&    -&    -&    -\\
$\Sigma(1706)1/2^+(P_{11})$&  1.6& 59.5& 28.3&    -& 10.3&    -&  0.4&  0.0&    -\\
$\Sigma(1607)3/2^-(D_{13})$&  0.3& 38.7& 49.0&    -& 12.0&  0.1&  0.0&  0.0&  0.0\\
$\Sigma(1669)3/2^-(D_{13})$& 12.1& 46.5&  5.8&    -& 30.9&  4.4&  0.1&  0.2&  0.1\\
$\Sigma(1767)5/2^-(D_{15})$& 40.2&  4.2& 24.4&    -& 30.9&  0.0&  0.0&  0.3&  0.0\\
$\Sigma(1890)5/2^+(F_{15})$&  3.6& 67.8& 12.7&  0.0& 11.2&  0.4&  0.1&  4.2&  0.0\\
$\Sigma(2025)7/2^+(F_{17})$& 26.9&  3.7&  8.0&  0.6&  3.0&  0.3& 15.4& 42.2&  0.0
\end{tabular}
\end{ruledtabular}
}
\end{table}
\begin{table}
\caption{\label{tab:br-b} 
Branching ratios for the decays of $\Lambda^*$ and $\Sigma^*$ resonances extracted from Model B.
Equations~(\ref{eq:br-bw})-(\ref{eq:gamma-quasi-kbsn}) are used for evaluating the ratios.
The quantum numbers for the $(\pi\Sigma^*)_i$ ($i=1,2$) and $(\bar K^* N)_i$ ($i=1,2,3$)
channels for a given $J^P$ are presented in Table~\ref{tab:qn}
}
{\footnotesize
\begin{ruledtabular}
\begin{tabular}{lrrrrrrrrr}
&\multicolumn{9}{c}{Branching ratios (\%)}\\
\cline{2-10}
Particle $J^P(l_{I2J})$ & 
$B_{\bar K N}$&
$B_{\pi \Sigma}$&
$B_{\eta \Lambda}$&
$B_{K \Xi}$&
$B_{(\pi\Sigma^*)_1}$&
$B_{(\pi\Sigma^*)_2}$&
$B_{(\bar K^* N)_1}$&
$B_{(\bar K^* N)_2}$&
$B_{(\bar K^* N)_3}$\\
\hline
$\Lambda(1512)1/2^-(S_{01})$& 63.6& 36.4&    -&    -&  0.0&    -&    -&    -&    -\\
$\Lambda(1667)1/2^-(S_{01})$& 36.3& 26.6& 21.2&    -& 15.5&    -&  0.3&  0.1&    -\\
$\Lambda(1548)1/2^+(P_{01})$& 24.0& 70.9&    -&    -&  5.0&    -&    -&    -&    -\\
$\Lambda(1841)1/2^+(P_{01})$& 23.7& 10.9&  4.0& 13.2& 14.9&    -&  0.5& 32.9&    -\\
$\Lambda(1671)3/2^+(P_{03})$&  3.9& 18.6& 43.2&    -& 33.9&  0.1&  0.2&  0.1&  0.0\\
$\Lambda(1517)3/2^-(D_{03})$& 43.1& 46.2&    -&    -& 10.1&  0.6&    -&    -&    -\\
$\Lambda(1697)3/2^-(D_{03})$& 31.8& 29.8&  0.1&    -&  2.9& 34.3&  0.0&  1.1&  0.0\\
$\Lambda(1924)5/2^-(D_{05})$&  5.4&  1.3&  0.1&  0.4& 85.7&  0.1&  2.4&  1.4&  3.1\\
$\Lambda(1821)5/2^+(F_{05})$& 56.5& 15.3&  0.5&  0.0& 25.2&  0.8&  0.0&  1.7&  0.0\\
$\Lambda(2041)7/2^+(F_{07})$&  1.6& 56.5&  1.7& 26.4&  9.5&  0.1&  1.5&  1.9&  0.9\\
\\
Particle $J^P(l_{I2J})$ &
$B_{\bar K N}$&
$B_{\pi \Sigma}$&
$B_{\pi \Lambda}$&
$B_{K \Xi}$&
$B_{(\pi\Sigma^*)_1}$&
$B_{(\pi\Sigma^*)_2}$&
$B_{(\bar K^* N)_1}$&
$B_{(\bar K^* N)_2}$&
$B_{(\bar K^* N)_3}$\\
\hline
$\Sigma(1551)1/2^-(S_{11})$& 45.6&  8.0& 46.3&    -&  0.0&    -&    -&    -&    -\\
$\Sigma(1940)1/2^-(S_{11})$& 53.4& 20.4&  1.9&  4.0&  3.3&    -&  3.2& 13.9&    -\\
$\Sigma(1457)1/2^+(P_{11})$&  1.2&  2.1& 96.7&    -&  0.0&    -&    -&    -&    -\\
$\Sigma(1605)1/2^+(P_{11})$&  3.6& 41.8& 43.4&    -& 11.2&    -&  0.0&  0.0&    -\\
$\Sigma(2014)1/2^+(P_{11})$& 10.4&  7.5& 10.0&  5.7& 17.2&    -& 16.1& 33.1&    -\\
$\Sigma(1492)3/2^-(D_{13})$&  0.0&  9.2& 90.7&    -&  0.1&  0.0&    -&    -&    -\\
$\Sigma(1672)3/2^-(D_{13})$&  6.8& 81.0&  6.6&    -&  1.6&  2.2&  0.0&  1.8&  0.0\\
$\Sigma(1765)5/2^-(D_{15})$& 39.7&  5.9& 18.3&    -& 36.1&  0.0&  0.1&  0.0&  0.0\\
$\Sigma(1695)5/2^+(F_{15})$&  0.3& 46.4& 53.0&    -&  0.2&  0.0&  0.0&  0.0&  0.0\\
$\Sigma(2014)7/2^+(F_{17})$& 29.1&  0.7&  6.3&  0.7& 57.8&  0.5&  1.0&  1.9&  2.0
\end{tabular}
\end{ruledtabular}
}
\end{table}

\section{$S$-wave $\Lambda$ resonances below the $\bar K N$ threshold}
\label{sec:below}

The nature of the $S$-wave ($J^P=1/2^-$) $\Lambda$ resonances lying below the $\bar K N$ threshold has 
long been an interesting subject since they are closely related to the extensively 
discussed $\Lambda(1405)$~\cite{l1405}.
In this section, we discuss such $S$-wave $\Lambda$ resonances extracted from our models.
However, here we add a caveat that Models A and B
were constructed by analyzing only the $K^- p$ reactions, and hence
the $\bar K N$ subthreshold region is beyond the scope of our current analysis.
Therefore, the results presented below should be considered as the ``predictions''
from our current models and are subject to change once our analysis is 
extended to include the data in the $\bar K N$ subthreshold region.
For this reason, we do not evaluate the uncertainties of the masses of these resonances.

\begin{figure}
\includegraphics[clip,width=0.5\textwidth]{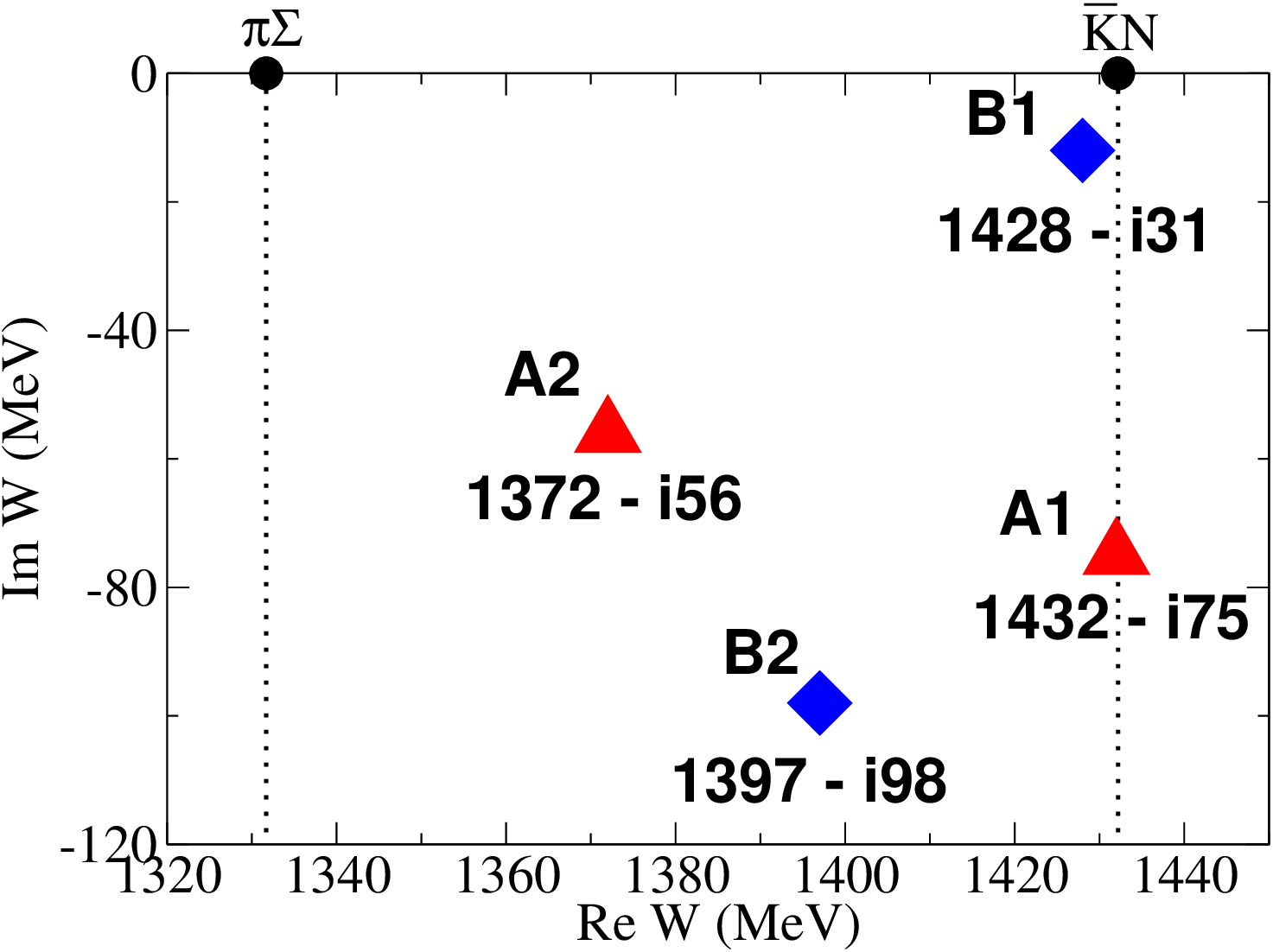}
\caption{(Color online)
$S$-wave ($J^P=1/2^-$) $\Lambda$ resonances in the $\bar K N$ subthreshold region.
Red triangles (blue diamonds) are resonance poles found from Model A (Model B).
}
\label{fig:s01-pole}
\end{figure}
Within our model, the 
$S$-wave $\Lambda$ resonances found in the region below the $\bar K N$ threshold
are presented in Fig.~\ref{fig:s01-pole}.
The red triangles and the blue diamonds represent the pole positions obtained 
from Model A and Model B, respectively. 
Both models predict two resonance poles for the $S$-wave $\Lambda$ 
resonances in the $\bar K N$ subthreshold region, while
their positions are rather different. 
This indicates a need of extending our  analysis to include the data in 
the $\bar K N$ subthreshold region.
Higher mass poles (A1 and B1 in Fig.~\ref{fig:s01-pole})
seem to correspond to $\Lambda(1405)$,
though the pole A1 has an imaginary part somewhat larger than 
what is usually expected for $\Lambda(1405)$.
The existence of another $\Lambda$ resonance with lower mass (A2 and B2 in Fig.~\ref{fig:s01-pole})
is similar to the result obtained in the so-called chiral unitary models (see, e.g., Ref.~\cite{ucm-1}).
In Table~\ref{tab:l-residue-2}, we list the residues for the $J^P=1/2^-$ $\Lambda$ resonances 
presented in Fig.~\ref{fig:s01-pole}.
\begin{table}
\caption{\label{tab:l-residue-2} 
Residues for $\pi \Sigma \to Y^* \to \pi\Sigma$ amplitudes [$R_{\pi \Sigma,\pi \Sigma}$ (MeV)] at 
the $J^P=1/2^-$ $\Lambda^\ast$ resonance pole positions found below $\bar K N$ threshold.
See the caption of Table~\ref{tab:a-residue-1} for the description of the table.
}
\begin{ruledtabular}
\begin{tabular}{llrr}
&Particle $J^P(l_{I2J})$ & 
\multicolumn{2}{c}{$R_{\pi\Sigma,\pi \Sigma}$}\\ 
\cline{3-4} 
&& $R$ & $\phi$\\
\hline
Model A
&$\Lambda(1372)1/2^-(S_{01})$&118&$-$68\\
&$\Lambda(1432)1/2^-(S_{01})$&177&  144\\
\\
Model B
&$\Lambda(1397)1/2^-(S_{01})$&142&$-$98\\
&$\Lambda(1428)1/2^-(S_{01})$&67& 110\\
\end{tabular}
\end{ruledtabular}
\end{table}

Since the poles $A1$ and $B1$ are located near the $\bar K N$ threshold,
it is interesting to examine the correlation between their pole values and 
the $S$-wave threshold parameters for the $I=0$ $\bar K N$ scattering amplitude,
as done for the $\Lambda(1670)1/2^-$ resonance near the $\eta \Lambda$ threshold 
(see Sec.~\ref{sec:res-mass}).
Similar to Eq.~(\ref{eq:eff-exp}), 
the $S_{01}$ $\bar K N$ scattering amplitude near the threshold can be written as
\begin{equation}
F^{I=0,S\mathrm{wave}}_{\bar K N,\bar K N}(k) \simeq k\times
\left(\frac{1}{a^{I=0}_{\bar KN}}-ik+\frac{r^{I=0}_{\bar KN}}{2}k^2 \right)^{-1} ,
\label{eq:eff-expkbn}
\end{equation}
where $a^{I=0}_{\bar KN}$ and $r^{I=0}_{\bar KN}$ are the scattering length and effective range
for the $I=0$ $\bar K N$ scattering, respectively.
These threshold parameters have been extracted in our previous paper~\cite{knlskp1}, 
and their values are:
\begin{equation}
a^{I=0}_{\bar KN} =
\left\{
\begin{array}{cc}
-1.37+i0.67~\mathrm{fm} & (\mathrm{Model~A}),\\
-1.62+i1.02~\mathrm{fm} & (\mathrm{Model~B}),
\end{array}
\right.
\end{equation}
\begin{equation}
r^{I=0}_{\bar KN} =
\left\{
\begin{array}{cc}
 0.67-i0.25~\mathrm{fm} & (\mathrm{Model~A}),\\
 0.74-i0.25~\mathrm{fm} & (\mathrm{Model~B}).
\end{array}
\right.
\end{equation}
By performing the same procedure as done for 
the $\Lambda(1670)1/2^-$ resonance in Sec.~\ref{sec:res-mass},
we find that the approximate amplitude~(\ref{eq:eff-expkbn})
has a pole at the complex $W$ with
\begin{equation}
W = E_{\bar K}(k) + E_N(k) =
\left\{
\begin{array}{cc}
1408-i56~\mathrm{MeV} & (\mathrm{Model~A}),\\
1427-i29~\mathrm{MeV} & (\mathrm{Model~B}).
\end{array}
\right.
\end{equation}
We find that the above value for Model B agrees well 
with the exact pole mass for the pole B1, while for Model A 
we find a significant deviation from the pole A1.
This can be understood because the position of the pole A1, as indicated in Fig.\ref{fig:s01-pole},
 is a bit far from 
the $\bar K N$ threshold and the approximated expression~(\ref{eq:eff-expkbn})
of the amplitude becomes less accurate than the case of the pole B1.
We also see in Fig.\ref{fig:s01-pole} that the poles $A2$ and $B2$ are even farther from the
$\bar K N$ threshold and hence they can not be well reproduced by the poles extracted from
the approximate
amplitude defined by Eq.~(\ref{eq:eff-expkbn}).

\section{Summary and future developments}
\label{sec:summary}

We have presented the parameters associated with the $\Lambda^*$ and $\Sigma^*$ resonances 
extracted from our DCC models that were constructed via a comprehensive partial-wave analysis of 
the $K^- p \to \bar K N, \pi\Sigma, \pi\Lambda, \eta\Lambda, K\Xi$ data~\cite{knlskp1}.
The extraction was accomplished by searching for poles of scattering amplitudes in the complex 
energy Riemann surface over the region with $m_{\bar K}+m_N < \mathrm{Re}(W) < 2.1$ GeV and 
$0<-2\mathrm{Im}(W) < 0.4$ GeV.
As a result, 18 (20) resonances are extracted from Model A (Model B) above the $\bar K N$ threshold.
The residues and branching ratios for the extracted resonances are also presented, and 
their values are found to be more sensitive to differences between the analyses
than the pole values.

Among the extracted resonances, a new narrow $J^P=3/2^+$ $\Lambda$ resonance
with $M_R=1671^{+2}_{-8}-i(5^{+11}_{-2})$~MeV is of particular interest.
Currently, this resonance is only found in Model B. 
However, the angular dependence of the
$K^- p \to \eta \Lambda$ differential cross section data seems to favor 
the existence of this resonance.
Given the fact that this new resonance is identified only through its contribution
to the  spin-averaged differential cross section
of the  $K^- p \to \eta \Lambda$ reaction
near the threshold, the 
 polarization data would be highly desirable to
have a definitive conclusion on the existence of  this resonance. 
Also, establishing low-lying $\Sigma^*$ resonances in 
$S_{11}$, $P_{11}$, and $D_{13}$ waves would also be an important task 
for the $Y^*$ spectroscopy.

By comparing the results from  our two models and
the KSU analysis, we found that the extracted resonance parameters have
significant analysis dependence.
This reflects the fact that the 
kinematical ($W$ and $\cos\theta$) coverage and accuracy of the available 
$K^- p$ reaction data are far from
``complete'' and not sufficient to eliminate analysis dependence on
the extracted resonance parameters.
More extensive and accurate data of the $K^- p$ reactions including 
the differential cross sections as well as the polarization observables 
(the recoil polarization $P$ and the spin-rotation parameters $\beta$, $R$, and $A$)
are definitely needed to get convergent results.
In fact, the data of all observables are relevant to accomplish 
an accurate extraction of amplitudes and resonance parameters with less analysis dependence,
as discussed in, for example, Ref.~\cite{shkl11}.
An impact of unmeasured observables of the $K^- p \to MB$ reactions 
for reducing analysis dependence has been explored in our previous paper~\cite{knlskp1}.
We have also found that the high-mass $\Lambda^*$ and $\Sigma^*$ resonances 
have large branching ratios to the quasi two-body $\pi \Sigma^*$ and $\bar K^* N$ channels,
suggesting that the data for the $2 \to 3$ reactions 
such as $K^- p \to \pi \pi \Lambda$ and $K^- p \to \pi \bar K N$
will also play an important role for establishing the high-mass resonances.
The experiments measuring these fundamental observables at 
the hadron beam facilities, such as J-PARC, will be essential
for making progress in establishing the $\Lambda^*$ and $\Sigma^*$ resonances.

As a byproduct of our $K^- p$ reaction analysis,
we have given ``predictions'' for the $J^P=1/2^-$ $\Lambda$ resonances located 
below the $\bar K N$ threshold. 
Both of our two models predict a resonance pole just below the $\bar K N$ threshold, 
which would correspond to $\Lambda(1405)1/2^-$.
Our two models also predict another $J^P= 1/2^-$ $\Lambda$ resonance pole 
with the mass $\sim$30-60~MeV lower than $\Lambda(1405)1/2^-$ we found.
This result is similar to what is obtained within 
the chiral unitary models~\cite{ucm-1}.
To make a decisive examination for the $J^P=1/2^-$ $\Lambda$ resonances, however, 
we must extend our analysis to include the data in the $\bar K N$ subthreshold region.
Our effort, in conjunction with the recent experimental initiatives~\cite{moriya,j-parc-e31}
will be published elsewhere.

\begin{acknowledgments}
This work was supported by the Japan Society for the Promotion of Science (JSPS) 
KAKENHI Grant No.~25800149 (H.K.) and 
Nos.~24540273 and~25105010 (T.S.),
and by the U.S. Department of Energy, Office of Nuclear Physics Division, 
under Contract No. DE-AC02-06CH11357.
H.K. acknowledges the support of the HPCI Strategic Program 
(Field 5 ``The Origin of Matter and the Universe'') of 
Ministry of Education, Culture, Sports, Science and Technology (MEXT) of Japan.
This research used resources of the National Energy Research Scientific Computing Center,
which is supported by the Office of Science of the U.S. Department of Energy
under Contract No. DE-AC02-05CH11231, and resources provided on Blues and/or Fusion,
high-performance computing cluster operated by the Laboratory Computing Resource Center
at Argonne National Laboratory. 
\end{acknowledgments}

\end{document}